\documentclass[reprint,amsmath,amssymb,aps,floatfix,superscriptaddress
]{revtex4-1}

\usepackage{graphicx}
\usepackage{dcolumn,xcolor}
\usepackage{bm}

\begin{document}


\title{The electrostatic charge on exuded liquid drops}

\author{Schuyler Arn}
\affiliation{Department of Physics, Emory University, 400 Dowman Dr., Atlanta, GA 30322, USA}
\author{Pablo Illing}
\affiliation{Department of Physics, Emory University, 400 Dowman Dr., Atlanta, GA 30322, USA}
\author{Joshua Mend\'{e}z Harper}
\email{joshua.mendez@pdx.edu}
\affiliation{Department of Electrical and Computer Engineering, Portland State University, Portland, OR 97201, USA}
\author{Justin C. Burton}
\email{justin.c.burton@emory.edu}
\affiliation{Department of Physics, Emory University, 400 Dowman Dr., Atlanta, GA 30322, USA}

\date{\today}

\begin{abstract}
Fluid triboelectrification, also known as flow electrification, remains an under-explored yet ubiquitous phenomenon with potential applications from material science to planetary evolution. Building upon previous efforts to position water within the triboelectric series, we investigate the charge on individual, millimetric water drops falling through air. Our experiments measured the charge and mass of each drop using a Faraday cup mounted on a mass balance, and connected to an electrometer. For pure water in a glass syringe with a grounded metal tip, we find the charge per drop ($\Delta q/\Delta m$) was approximately -5 pC/g to -1 pC/g. This was independent of the release height of the drop, tip diameter and length, tip cleaning preparation, and whether the experiment was shielded with a Faraday cage. Biasing the tip to different voltages allowed for linear control of the drop charge, and the results were consistent with known electrochemical effects, namely the Volta potential expected between most metals and bulk water ($\approx$ -0.5 V). Introducing insulating plastic materials into the experiment (from the syringe body or tip) imparted large amounts of charge on the drops with systematic charge evolution. Together these results show that the flow electrification of water is more complex than previously reported, and is driven by material-dependent electrostatic processes.  
\end{abstract}

\maketitle

\section{Introduction}
Triboelectric charging--encompassing both frictional and contact electrification--is so commonplace that we often overlook its mysterious nature. The shock that one feels when touching a doorknob, the electric crackling of clothes as these are pulled from a mechanical laundry dryer, or even the whimsical play between one's hair and a latex balloon are all underpinned by charge transfer between materials during contact. In nature, triboelectricity manifests in varied geophysical contexts, from the dramatic volcanic lightning on Earth, \cite{mcnutt2010volcanic,mendez2016effects} to electrified dust storms and dunes on other worlds \cite{mendez2017electrification, mendez2021detection}. Despite our familiarity with triboelectrification (simplistically referred to as ``static"), the mechanisms driving charge exchange between two surfaces remain imperfectly understood.  
Evidently, relative motion (rubbing) between the surfaces is important, suggesting that sharp and large relative shear effectively drive charge transfer \cite{gorman2024electrostatic, james2000volcanic}. However, frictional interactions cannot alone account for contact electrification, since significant charging can occur even if surfaces lightly osculate \cite{apodaca2010contact}. 

Whether triboelectric charging results from bulk material transfer, electron or ion transfer, some interfacial adsorbate, or from a combination of all these remains unclear \cite{Lacks2019}.  During the last few decades, however, evidence has emerged that water plays a key role in generating charge separation\cite{harris2019temperature,Lacks2019, grosjean2023single}, stabilizing existing interfacial charge \cite{baytekin2011water}, and dissipating charge \cite{mendez2022lifetime, mendez2024strategies, mendez2024moisture}. A salient role of water in triboelectrification is perhaps unsurprising: materials in ambient air typically host layers of adsorbed water on their surfaces, unless they are particularly hydrophobic.  Dissociated H$^+$ and OH$^-$ ions within these layers may allow for charge exchange during collisions or frictional interactions \cite{gu2013role, zhang2015electric}. That water layers modulate charge exchange in systems of solids implies that triboelectrification may also occur at fluid-solid interfaces (or even fluid-fluid interfaces!)\cite{wang2021contact,jin2022electrification, armiento2022liquid}. As in solid-solid triboelectrification, the chemical composition of the liquid-solid pair tunes the character of the charging. If the fluid moves relative to the solid, charge separation is known as ``flow electrification'' \cite{TOUCHARD2001}. For viscous fluids, the no-slip assumption implies that the shear forces at the fluid-solid boundary are smaller than those between two frictional solids. \textcolor{black}{However, the contact area is orders of magnitude larger than expected for solid-solid frictional contacts \cite{rubinstein2006contact} since the liquid coats the entirety of the solid surface.} Even in the absence of flow, there can be a static potential difference between a bulk solid and liquid in contact due to differences in electronic energy states \cite{li2021linear}.

Flow electrification has been examined for more than a century, primarily by the petroleum and lubrication engineering industries \cite{Klinkenberg1958,Solomon1959,ASTM1991,Harvey2002,TOUCHARD2001}. Within these contexts, electrostatic charging and discharge pose a significant hazard in dielectric and combustible hydrocarbon fuels \cite{mackeown1942electrical,Bustin1983,Yuqin2013}. Consider that more than 30 unintended fuel-air mixture ignitions during vehicle refueling were reported in Germany alone between 1992 and 1995. These incidents were ultimately attributed to electrostatic accumulation and subsequent discharge of flowing liquids\cite{von1997avoidance}. Charging in dielectric fluids like hydrocarbons is particularly hazardous in filtration systems \cite{Gavis1968,Huber1977} and pipe flow\cite{BOGRACHEV2012,Abedian1982,CABALEIRO2008} where the contact area between the liquid and wall is large.  For insulating fluids, the Debye layer thickness can extend well beyond the flow boundary layer (more than a few microns). As such, ions in the Debye layer are easily entrained in flow. Additionally, because these fluids have poor conductivities, they do not readily allow for charge recombination. Although this current is often treated phenomenologically, very few studies consider the underlying mechanism of flow electrification \cite{OSTERN1924,eladawy2011,Touchard1996,Washabaugh1997}.

\textcolor{black}{Flow electrification also occurs in liquids with higher conductivities like water. In 1892, inspired by observations of negative charge developed in air around waterfalls across the Alps, Lenard conducted a series of laboratory experiments to study the charging behavior of falling water drops or streams in ambient air \cite{lenard1892ueber}. These pioneering experiments showed that the water dispensed from a reservoir collects charge while the surrounding air develops an oppositely-charged potential. Lenard noted variations in this behavior depending on the cleanliness and source of the dispensed water, among other material properties used in the experiments.}
The Debye length in pure water is $\sim$0.7 nm, orders of magnitude smaller than in dielectric liquids with very few ions. This has led to proposals to use water as a power source for nanoscopic graphene-based devices \cite{Zhang2021}. Another recent study by \citeauthor{BURGO2016} presented a triboelectric series for water flowing against materials ranging from air to copper to polytetrafluoroethylene, or PTFE. Both the magnitude and the sign of the flow electrification of water was highly dependent on the composition of the other material \cite{BURGO2016}. 

Interestingly, \citeauthor{BURGO2016} showed that water charged positively against all materials with the exception of air. The fact that air flow can reliably charge water and potentially other liquids is surprising since ambient air contains only $\sim$1 ion/mm$^3$ \cite{mendez2022lifetime}. Furthermore, under fair weather conditions, near-surface air contains slightly more positive ions than negative ones \cite{kamra1982fair}, suggesting that any falling drop would scavenge a net positive charge. In other words, merely collecting ions from the air is insufficient to explain the magnitude and polarity of observed flow electrification between water and air by Burgo and coworkers. Such incongruities, particularly in the context of a ubiquitous material like water, underscore the need for more detailed investigations into contact electrification across fluid-solid and fluid-fluid interfaces.

Here, we report on experiments designed to characterize the charge on individual deionized water drops falling from glass syringes with a variety of needles. The drops, which grow quasi-statically from the tip of the needle, fall into a custom-made Faraday cup, and the mass of each drop is simultaneously measured with a precision balance. In the absence of plastics, we find little or no flow electrification from the ambient environment--that is, the drop charge does not depend on deposition height, metallic tip length, flow rate, or the presence of electrostatic shielding surrounding the experiment. The charge per unit mass on each drop, $\Delta q/\Delta m$, ranged from -5 to -1~pC/g, consistent with the expected Volta potential between typical metals and bulk water \cite{li2021linear}. Adding salt to the water reduces the overall charge by $\sim$50\%. In contrast, introducing common plastic components in the experimental setup (such as a polypropylene syringe body or PTFE syringe tip) can drastically change the charge on each drop. In the case of a PTFE tip, for instance, the absolute charge-to-mass ratio can be as large as 120~pC/g. Furthermore, the charge polarity gained by drops flowing against plastics depends on the material history (e.g. how long fluid has been flowing through a plastic tip) and environmental conditions. Our results suggest that the electrification of individual water drops flowing against metals may be understood from known electrochemical effects. However, interactions with plastics can give rise to more complex charge exchanges whose underlying mechanisms require further clarification.  

\begin{figure*}[t]
    \centering
    \includegraphics[width = \textwidth]{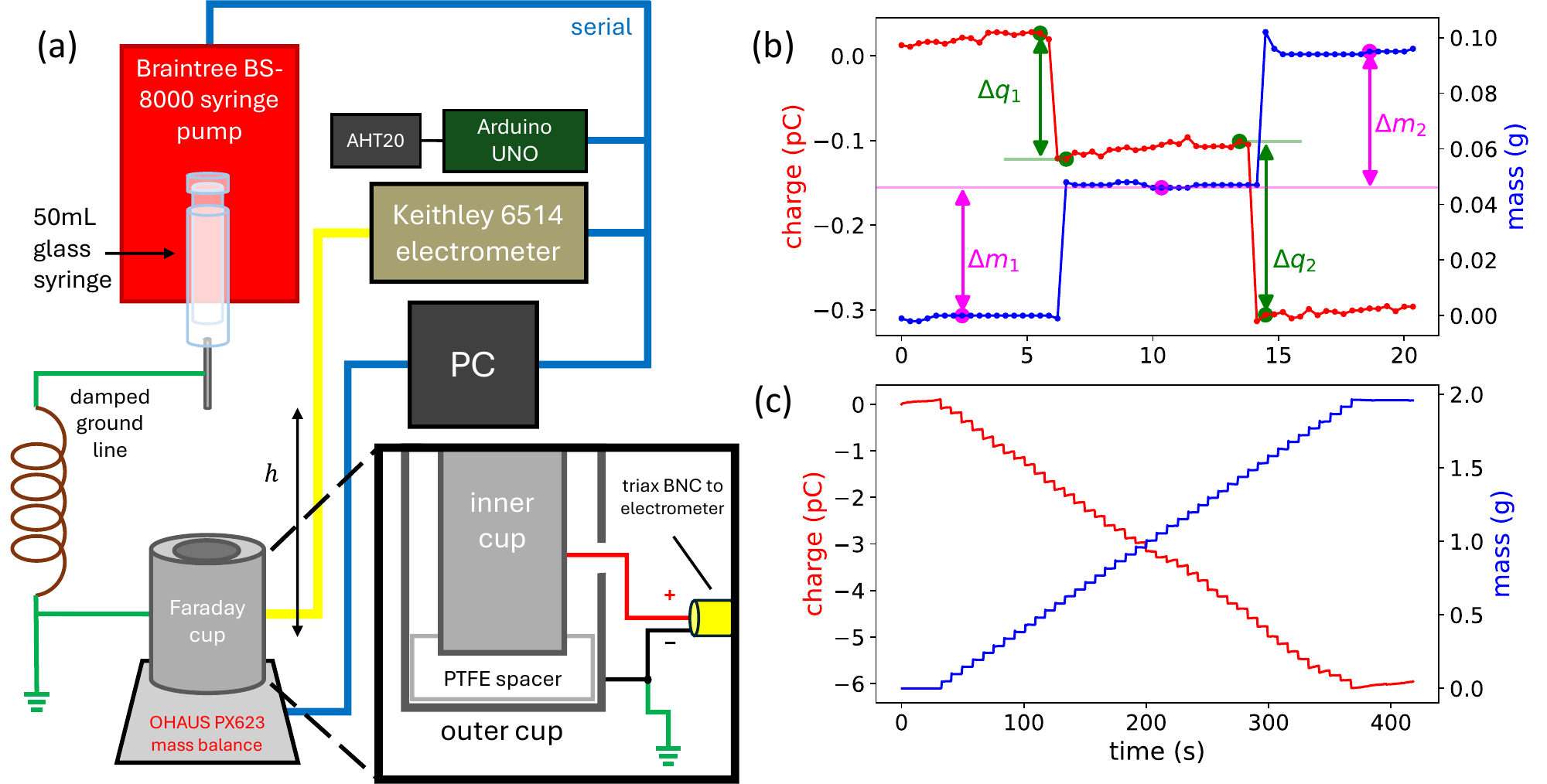}
    \caption{Schematic of the experimental setup and sample data for measuring charge on individual drops. (a) Wiring diagram depicting the connection of the Faraday cup (FC) to the electrometer for charge measurements. (b) Experimental data for two sequential DI water drops deposited from a 50~mL glass syringe connected to a stainless steel tip of length 5~cm and 2.39~mm ID, recorded at 44\% RH. There are distinct jumps in both mass and charge. Mass plateau medians are indicated by magenta markers, and their difference provides $\Delta m$ for each drop, such as the labeled $\Delta m_{1}$ and $\Delta m_{2}$. Green markers indicate the points on either side of the charge jumps used to calculate $\Delta q$ for each drop, namely $\Delta q_{1}$ and $\Delta q_{2}$. (c) Entire time series showing the deposition of 2~mL of water under the same experimental conditions as in (b). For this experiment, $\langle \Delta q \rangle =$ -0.17 $\pm$ 0.03~pC, and $\langle \Delta m \rangle =$ 0.048 $\pm$ 0.002~g.}
    \label{exp_setup}
\end{figure*}

\section*{Experimental Procedure}

To investigate the charging of drops, we employed the setup depicted in Fig.~\ref{exp_setup}a. A programmable syringe pump (Braintree Scientific BS-8000) was fixed vertically above a custom-made Faraday cup (FC) capable of handling liquids. We programmed the pump to dispense individual drops from a luer-lock syringe into the FC at regular intervals. \textcolor{black}{The FC consisted of an inner sensing cup nested within a grounded shielding cylinder separated by a PTFE spacers. The inner cup was electrically connected to the center pin a triax bulkhead connector, while the outer housing was grounded. The inner cup had a diameter of 4.57 cm and a depth of 3.05 cm, yielding a total internal volume of 50 mL. Individual FC components were ultrasonically cleaned, dried in an oven, and allowed to cool before assembly and employment in our experimental setup.  An electrometer (Keithley 6514) connected to the FC using a flexible, low-noise triax cable (operating in the <20 nC range) allowed us to measure charge on falling drops with a resolution of 0.01 pC. } 

\textcolor{black}{To measure the mass of individual drops, the FC was} placed on a precision balance (Ohaus PX623) with milligram resolution. \textcolor{black}{Thus, we were able to measure the charge-to-mass} ratio of each drop, $\Delta q/\Delta m$, and compare our data with that of previous investigations of fluid electrification \cite{BURGO2016, hendricks1962charged, snarski1991experiments}. Lastly, we monitored the ambient conditions using a temperature and humidity sensor (Aosong AHT20). For all experiments, temperature and relative humidity (RH) varied in the range of 20--22$^{\circ}$C and 20--50\%, respectively. The four variables--charge, mass, temperature, and RH--were sampled every $\sim$0.3~s across the duration of an experiment.

Depending on the inner diameter (ID) of a syringe tip (1.35--2.69~mm) and the flow rate of the syringe pump, drops were produced with a period of 5--10~s and \textcolor{black}{radius of 1.85--2.35~mm. For most trials, we employed a flow rate of 350 $\mu$L/min, but also tested a range of flow rates (35--750 $\mu$L/min) were individual drops could be resolved by the mass balance. These results are shown in Fig.~\ref{exp_setup}e-f, and show little variation in drop charge and mass with the flow rate}. Drops were allowed to fall 10--40~cm in air before landing in the FC. For some experiments, we minimized the influence of spurious electric fields by enclosing the entire experimental setup in a grounded Faraday cage made from aluminum wire mesh. Lastly, any metallic parts of the syringe (e.g. the tip or the metallic base of the syringe luer-lock system) were grounded to the outer shell of the FC using a small-gauge, coiled, solid-core copper wire. The solid core coil minimized any mechanical coupling between the syringe and the FC that would cause errors in mass measurements.


The charge $q$ on a drop entering the FC is registered as a step change in the voltage $V$ across the known feedback capacitor $C$ of the electrometer, $V =q/C$. Similarly, the addition of a drop into the FC causes a stepwise jump in the mass reading. Typical recordings of charge (red curve) and mass (blue curve) for two consecutive drops are shown in Fig.~\ref{exp_setup}b. A sudden change in the balance reading indicates an impinging drop. \textcolor{black}{To allow for stabilization after impact,} we measured the center point on the plateau, $m_i$, between two consecutive impinging drops, and estimated the mass of an individual drop $\Delta m$ by computing the difference in readings between two neighboring plateaus $\Delta m= m_i - m_{i-1}$. We used a similar procedure to extract the charge on a falling drop: a sudden change in the electrometer output corresponds to a drop entering the FC. However, electrometers of this sensitivity are prone to drift across the measurement timescales. Thus, we calculated the charge on a drop ($\Delta q$) by subtracting the electrometer reading immediately before and after the arrival of the drop (as detected by the mass change). A zoomed-out time series showing the mass and charge evolution for 41 consecutive drops is shown in Fig.~\ref{exp_setup}c.

Most experiments were conducted with deionized, ultra pure, filtered water (ELGA Veolia Purelab Chorus 1 Reservoir). The water was handled in clean glass containers before being deposited into the testing syringe. However, we also performed experiments with salt water, glycerol, and Fluorinert FC-70. Pure NaCl and glycerol were obtained from Sigma Aldrich, and the Fluorinert was obtained from 3M. To explore the dependence of solid surface properties on the electrification of fluids, we dispensed individual drops from syringe bodies and tips of varied materials. Most experiments used an all-glass, 50~mL capacity syringe with a stainless steel luer-lock mounted at the base (Tomopal). The syringe was cleaned with pure ethanol, rinsed with deionized water, and then dried in an oven at 50$^{\circ}$C prior to use. The glass syringe was connected to various grounded 304 stainless steel or custom nickel-plated stainless steel syringe tips. For nearly all experiments, the tips were cleaned with ethanol and rinsed with water prior to use, but we also tried two other methods of cleaning syringe tips: rinsing them with Neutrad solution and exposing them to oxygen plasma with a custom-built oxygen plasma oven \cite{mouat2020tuning,pye2018precursors}. In addition to glass and metals, we conducted experiments with plastic components since these are commonly used in fluid electrification experiments \cite{BURGO2016} and broad fluid transport applications. In one set of experiments, we used 20~mL polypropylene syringes (Henke-Ject) connected to standard luer-lock syringe tips composed of a polypropylene luer base and a 304 stainless steel, flat-ended cylindrical tube (1.70~mm ID). In another set of experiments, we used the glass syringes, but all-plastic PTFE syringe tips (1.35~mm ID). All metal and plastic syringe tips were obtained from Vita Needle.

\section*{Results and Discussion}

\begin{figure}[!]
    \centering
    \includegraphics[width = \columnwidth]{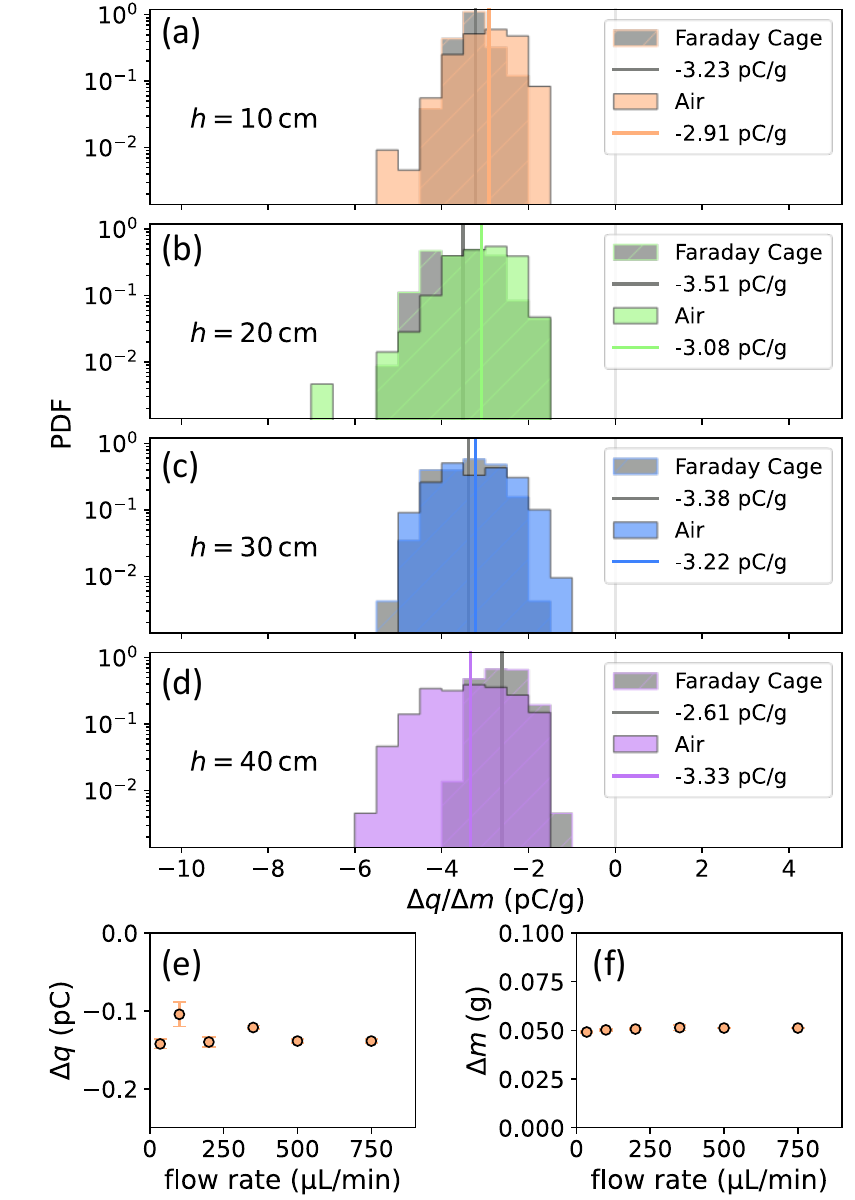}
    \caption{Probability distribution functions (PDF) for $\Delta q/\Delta m$ using a glass syringe body connected to a 5 cm long, 2.39 mm ID syringe tip. The flow rate was 350 $\mu$L/min, and data was recorded at 44--45\% RH. Each distribution is computed over $n \geq$ 410 data points. Each panel (a-d) represents a different deposition height ($h$), and also includes data taken with a Faraday cage enclosing the syringe pump, cup, and balance assembly (Fig.~\ref{exp_setup}a). The solid lines indicate the mean of each distribution, and the variance in the data primarily comes from the variance in $\Delta q$ since the drop mass was fairly uniform. For completeness, for the $h=$ 10 cm distributions, we found $\langle \Delta q_\mathrm{air} \rangle =$ -0.11 $\pm$ 0.02 pC, $\langle \Delta q_\mathrm{cage} \rangle =$ -0.13 $\pm$ 0.01 pC, $\langle \Delta m_\mathrm{air} \rangle =$ 0.037 $\pm$ 0.006 g, and $\langle \Delta m_\mathrm{cage} \rangle =$ 0.042 $\pm$ 0.005 g. For the $h=$ 20 cm distributions, we found $\langle \Delta q_\mathrm{air} \rangle =$ -0.13 $\pm$ 0.04 pC, $\langle \Delta q_\mathrm{cage} \rangle =$ -0.14 $\pm$ 0.02 pC, $\langle \Delta m_\mathrm{air} \rangle =$ 0.042 $\pm$ 0.009 g, and $\langle \Delta m_\mathrm{cage} \rangle =$ 0.040 $\pm$ 0.004 g. For the $h=$ 30 cm distributions, we found $\langle \Delta q_\mathrm{air} \rangle =$ -0.15 $\pm$ 0.04 pC, $\langle \Delta q_\mathrm{cage} \rangle =$ -0.15 $\pm$ 0.03 pC, $\langle \Delta m_\mathrm{air} \rangle =$ 0.047 $\pm$ 0.005 g, and $\langle \Delta m_\mathrm{cage} \rangle =$ 0.046 $\pm$ 0.003 g. Lastly, for the $h=$ 40 cm distributions, we found $\langle \Delta q_\mathrm{air} \rangle =$ -0.14 $\pm$ 0.04 pC, $\langle \Delta q_{cage} \rangle =$ -0.11 $\pm$ 0.02 pC, $\langle \Delta m_\mathrm{air} \rangle =$ 0.041 $\pm$ 0.005 g, and $\langle \Delta m_\mathrm{cage} \rangle =$ 0.041 $\pm$ 0.003 g. \textcolor{black}{(e) Plot of $\langle \Delta q \rangle$ against flow rate. (f) Plot of $\langle \Delta m \rangle$ against flow rate. Error bars were produced from the standard deviation of per-trial means at each flow rate, and all trials were completed at $h =$ 10 cm.}}
    \label{heights_pdf}
\end{figure}

\begin{figure}[t]
    \centering
    \includegraphics[width = \columnwidth]{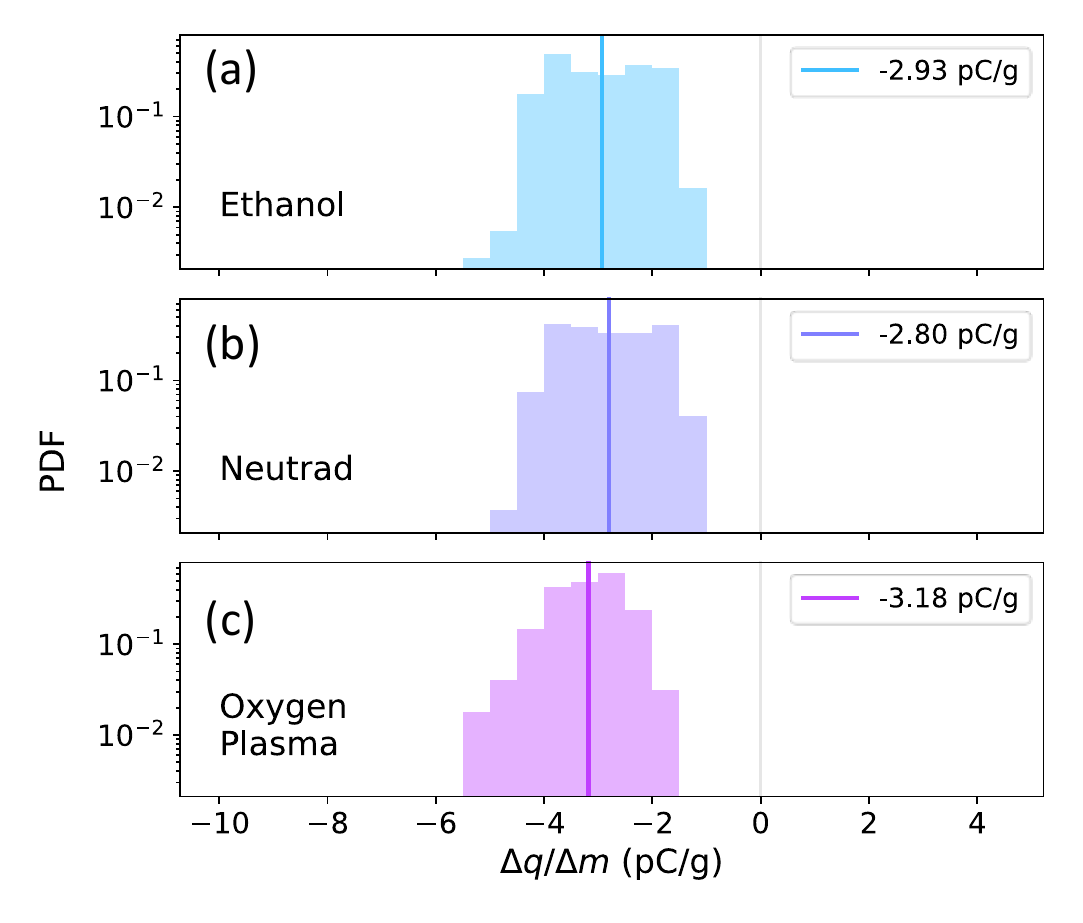}
    \caption{
    PDFs for $\Delta q/\Delta m$ under different cleaning procedures. Experiments used a glass syringe body connected to a 5 cm long, 2.39 or 2.69 mm ID syringe tip at a flow rate of 350 $\mu$L/min with 2~mL total volume. Each distribution is computed over $n \geq$ 450 data points. (a) The 2.39 mm ID tip was rinsed with pure ethanol followed by multiple rinses with DI water, yielding $\langle \Delta q \rangle =$ -0.12 $\pm$ 0.03 pC and $\langle \Delta m \rangle =$ 0.043 $\pm$ 0.002 g. Trials were recorded at 34\% RH. (b) The 2.39 mm ID tip was placed in an ultrasonic bath of 2\% w/w Neutrad solution for 30 minutes followed by multiple rinses with DI water, yielding $\langle \Delta q \rangle =$ -0.12 $\pm$ 0.03 pC and $\langle \Delta m \rangle =$ 0.044 $\pm$ 0.003 g. Trials were recorded at 20\% RH. (c) The larger, 2.69 mm ID tip was cleaned in an oxygen plasma cleaner at a pressure of 500 mTorr for 30 s and then allowed to cool, yielding $\langle \Delta q \rangle =$ -0.11 $\pm$ 0.02 pC and $\langle \Delta m \rangle =$ 0.034 $\pm$ 0.007 g. Trials were recorded at 48\% RH. Since the charging behavior showed little to no variation between cleaning procedures, the ethanol cleaning procedure was used unless otherwise stated.  }
    \label{cleaning_pdf}
\end{figure}

The aggregated results of our experiments with all glass and metal syringes are illustrated in Fig.~\ref{heights_pdf}a-d, which shows histograms of $\Delta q/\Delta m$ for many experimental conditions. We found little to no dependence of $\Delta q/\Delta m$ on the drop deposition height ($h$) or the presence of a Faraday cage shielding the experimental apparatus. The distributions were assembled from multiple independent experiments using 2~mL of water and were completed over multiple days. The average value of $\Delta q/\Delta m$ for each distribution is shown as a vertical line. Most drops fall within the range of -2.0~pC/g to -4.5~pC/g (5th and 95th percentile, respectively). The addition of a protective Faraday cage around the entire setup made a noticeable difference for $h=$ 40~cm, but the shift of the distribution mean was smaller than the standard deviation. Taken together, these results suggest that the characteristic charge on each drop is determined by the materials in contact and its history, and not its interaction with air as it falls from the syringe needle.

As noted above, we cleaned the syringe needles using three different preparation methods: rinsing them with pure ethanol followed by DI water, soaking them in a Neutrad ultrasonic bath then rinsing with DI water, and exposing them to an oxygen plasma for 30~s. The effects of different cleaning procedures are summarized in Fig.~\ref{cleaning_pdf}. The histograms of $\Delta q/\Delta m$ suggest that the charge gained by water droplets in contact with metal are independent of cleaning procedure. We note, however, that the oxygen plasma cleaning did decrease average drop mass, presumably because it changed the contact angle of the water and stainless steel contact line. This affects the formation of the drop near the attachment to the syringe tip, and ultimately the force balance that determines the maximum weight of the drop that can be supported.  


\begin{figure}
    \centering
    \includegraphics[width = \columnwidth]{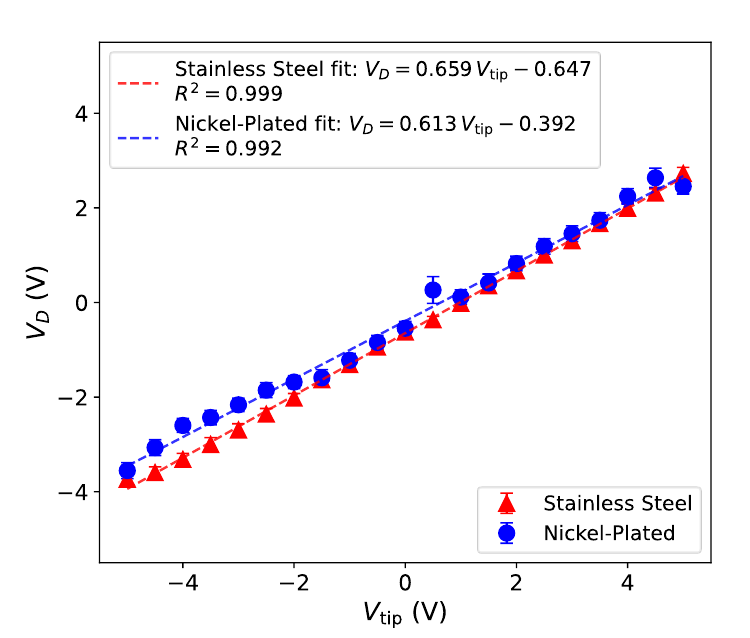}
    \caption{Bias voltage applied to a 5 cm long, 2.69 mm ID syringe tip ($V_\text{tip}$) versus the average voltage ($V_D$) of each drop, calculated using Eq.~\ref{eq_sphcap}. 
    Error bars were produced from the standard deviation of $n>$ 110 individual drops. Dashed lines depict linear fits to the data and are indicated in the legend. Data for the stainless steel tip and the nickel-plated tip were acquired at 49\% and 34.5\% RH, respectively.}
    \label{voltage_bias}
\end{figure}

\subsection*{Volta potential}

Having measured charge and mass independently, we can estimate the voltage, $V_D$, on each drop of radius $r$ by assuming they are spherical capacitors:
\begin{equation}
    V_D=\dfrac{Q}{4 \pi \epsilon_0 r} = \dfrac{\Delta q}{4 \pi \epsilon_0}\left(\dfrac{4 \pi \rho_w}{ 3 \Delta m}\right)^{1/3}.
    \label{eq_sphcap}
\end{equation}
Above, $\rho_w$ is the density of pure water, and $\epsilon_0$ is the permittivity of free space. For the typical values of $\Delta q$ and $\Delta m$ in our experiments, $V_D=$ -0.3~V to -0.8~V. The invariance of these measurements suggests that the potential difference between the grounded metallic syringe tip and the water drops is an inherent property of the metal-water interface. In fact, these voltage values are consistent with Volta potentials, $\Delta\psi$, for metallic surfaces in contact with bulk water \cite{li2021linear,le2017PRL}. Typically, Volta potentials vary from -0.3~V to -1.0~V for $sp$ and transition metals, with a dependence on the exposed crystalline structure \cite{li2021linear}. 

The Volta potential is often measured with scanning Kelvin probe force microscopy \cite{ma2021skpfm, guo2012water, ornek2019volta}, and is defined as the difference between the potential of zero charge, $U_{pzc}$, and the work function of the metal, $\Phi$:
\begin{equation}
    \Delta\psi = eU_{pzc} - \Phi = -\delta\chi^M_0 + g^\text{solv}(\text{dip})_0.
    \label{volta}
\end{equation}
The potential of zero charge is analogous to the work function when the metal is in contact with a solution instead of vacuum. The Volta potential can be thought of as the bulk potential of a solution (relative to vacuum) when in contact with a metal surface. From theory, $\Delta\psi$ has two contributions: a reorientation of water molecules at the metal surface, $g^\text{solv}(\text{dip})_0$, and a redistribution of surface metal electrons, $-\delta\chi^M_0$ \cite{trasatti1999potential, mohandas2024understanding}. To further investigate the possibility that $V_D$ is a natural electrochemical bias developed between the metal interface and the bulk water, we used a power supply to bias the voltage of our metallic syringe tips (both bare and nickel-plated 304 stainless steel) during drop deposition into the FC. The results are illustrated in Fig.~\ref{voltage_bias}. The voltage of the syringe tip, $V_{\text{tip}}$, was measured relative to the outer grounded shell of the FC, and $V_D$ was calculated from Eq.~\ref{eq_sphcap}. 

\begin{figure}[t]
    \centering
    \includegraphics[width = 0.75\columnwidth]{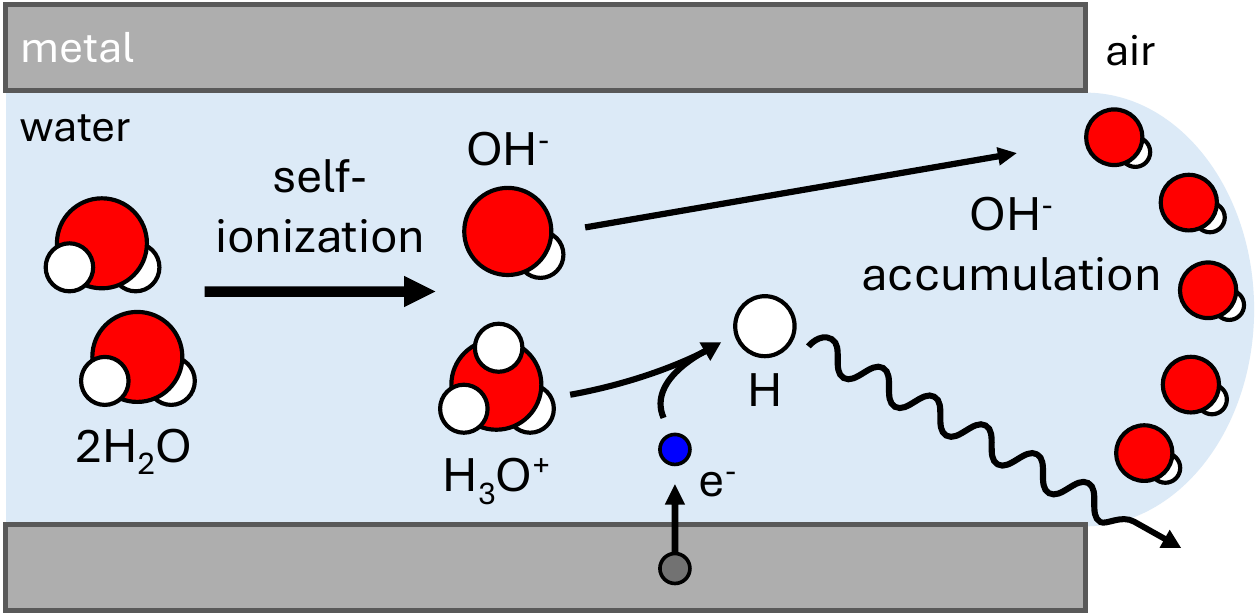}
    \caption{\textcolor{black}{Diagram of a potential charging mechanism for pure water in contact with a conducting metal surface. Paired water molecules self-ionize into hydroxide (OH$\textsuperscript{-}$) and hydronium (H$\textsubscript{3}$O$\textsuperscript{+}$) ions. A small fraction (less than 1 ppm) of H$\textsubscript{3}$O$\textsuperscript{+}$ ions source an electron (e$\textsuperscript{-}$) from the metal, and ultimately produce neutral hydrogen gas (H). This leads to a net charge imbalance where the hydroxide ions preferentially accumulate at the air-water interface of the drop. This process continues until the drop has reached its equilibrium Volta potential, $\Delta\psi$, between the water and metal. The neutral hydrogen will eventually form diatomic hydrogen (H\textsubscript{2}) and escape the drop into ambient air.}}
    \label{water_ions}
\end{figure}

The data for both syringe tips can be fit to a linear relationship. The intercept with the vertical axis ($V_{\text{tip}}=$ 0) agrees well with our measurements using a grounded tip (Fig.~\ref{heights_pdf})--that is, it represents the Volta potential, $\Delta\psi$. Interestingly, the slope of the linear fit is not unity: a small increase in $V_\text{tip}$ leads to a slightly smaller increase in $V_D$. This result may be expected since the applied voltage can change the distribution of electrons at the surface and the water dipole contribution to the Volta potential in Eq.~\ref{volta}~\cite{mohandas2024understanding}. Although we did not perform independent measurements of $\Delta \psi$ using alternative experimental techniques~\cite{le2017PRL}, the consistency of $V_D$ over multiple experimental conditions and its quantitative agreement with the expected values of $\Delta\psi$ strongly suggest their equivalency. 

\textcolor{black}{In order to obtain a potential difference between the interior bulk water and the ambient air, there must be a net surface charge density at the air-water interface. In water, hydroxide ions (OH$\textsuperscript{-}$) and hydronium ions (H$\textsubscript{3}$O$\textsuperscript{+}$) are expected to accumulate at the air-water interface \cite{creux2007specific}. A net negative charge can be due to an imbalance in the natural ion concentration. In pure water, the concentration of each ion species is approximately 1.7 parts per billion (ppb). For a water drop of radius 2 mm, this corresponds to 1.9$\times$10$^{12}$ ions of each species. Additionally, the charge imbalance required to produce a potential $V_D$ = -0.5 V is about 7$\times$10$^5$ elementary charges. This means that only a small fraction (less than 1 ppm) of the naturally occurring ions would need to contribute to the net charge in order to produce the expected Volta potential. A potential mechanism to produce this charge imbalance is shown in Fig.~\ref{water_ions}. A hydronium ion can source an electron from the metal surface, eventually resulting in a minute amount of hydrogen gas that either remains dissolved or escapes into the air. We note that this charge imbalance would produce negligible changes in pH, and thus absorbed gases, such as CO$\textsubscript{2}$, could also produce a similar charge imbalance.}




\begin{figure}[t]
    \centering
    \includegraphics[width = \columnwidth]{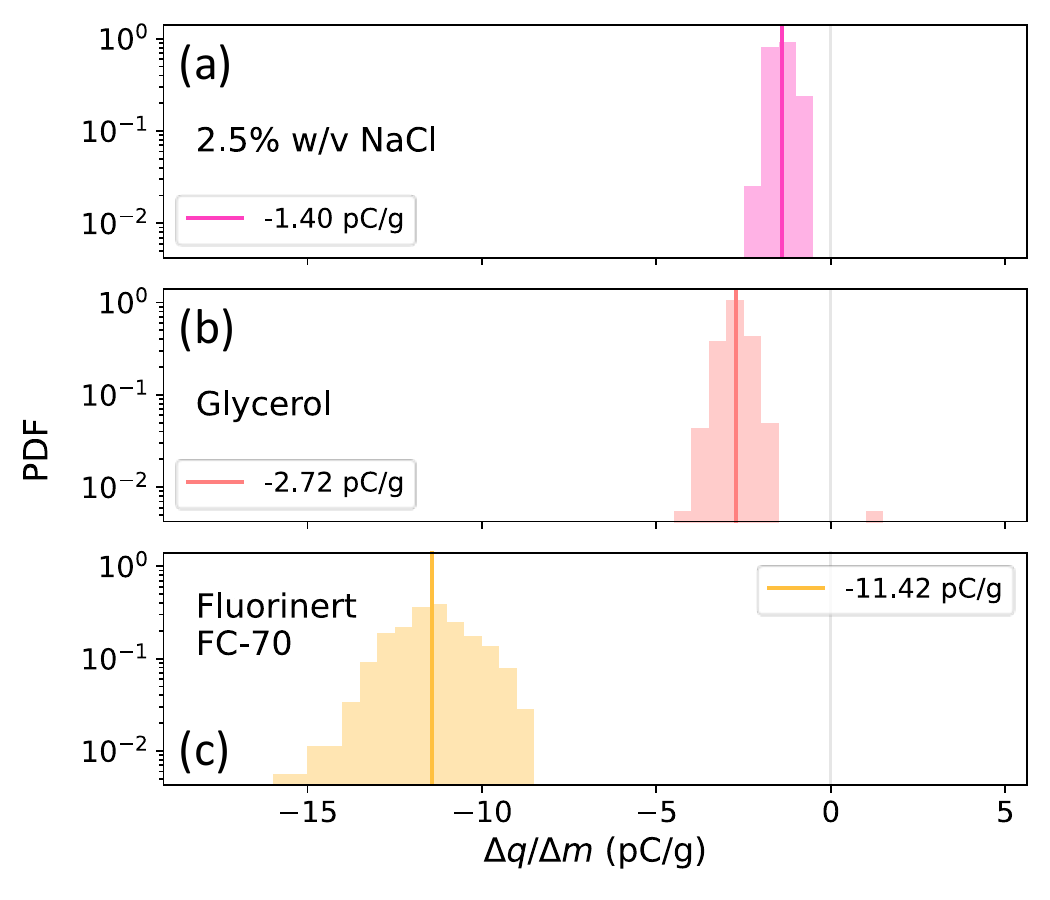}
    \caption{PDFs of $\Delta q/\Delta m$ for different fluids dispensed from a 5 cm long, 1.60 or 2.39~mm ID stainless steel luer lock syringe tip. The salt solution and glycerol data were acquired at a flow rate of 350 $\mu$L/min on the 2.39~mm ID tip, while the high density and much lower surface tension of Fluorinert FC-70 required a flow rate of 35 $\mu$L/min on the 1.60~mm ID tip. Each distribution is computed over $n>$ 360 data points. (a) For salt water, $\langle \Delta q \rangle =$ -0.08 $\pm$ 0.02 pC and $\langle \Delta m \rangle =$ 0.054 $\pm$ 0.004 g. Data were captured at 48\% RH. (b) For glycerol, $\langle \Delta q \rangle =$ -0.13 $\pm$ 0.02 pC and $\langle \Delta m \rangle =$ 0.049 $\pm$ 0.002 g. Data were captured at 48\% RH. (c) For Fluorinert FC-70, \textcolor{black}{$\langle \Delta q \rangle =$ -0.18 $\pm$ 0.02 pC and $\langle \Delta m \rangle =$ 0.016 $\pm$ 0.001 g}. Data were captured at 29\% RH. Overall, the addition of ions (salt) in the fluid narrows the distribution, whereas non-polar and highly insulating fluids display broader distributions.}
    \label{fluids_pdf}
\end{figure}

\subsection*{Insulating fluids}

{\color{black}The mechanism depicted in Fig.~\ref{water_ions} results from the naturally occurring ions and concomitant sub-nanometer Debye length in water. However, the charging mechanisms at play in insulating fluids may be more complex.} Thus, using the same glass syringe and stainless steel tips, we also investigated other liquids with the same procedure as for pure water. Figure~\ref{fluids_pdf} shows histograms of $\Delta q/\Delta m$ for a 2.5\% w/v NaCl aqueous solution, pure glycerol, and a fluorinated hydrocarbon fluid (Fluorinert FC-70). For salt water, both the magnitude and standard deviation of $\Delta q/\Delta m$ were slightly smaller than those of DI water. The presence of free ions in the solution can reduce the potential of zero charge (making it closer to the metal work function), and thus reduces the magnitude of the Volta potential \cite{ornek2019volta}. \textcolor{black}{Moreover, aqueous ions such as Na$\textsuperscript{+}$ and Cl$\textsuperscript{-}$ are expected to be less important for interfacial charging mechanisms due to their lack of hydrogen bonding~\cite{zimmermann2001electrokinetic}.} For glycerol, drop radii ranged from 2--2.2 mm, and $\Delta q/\Delta m$ was close to that of pure water. Although we do not have a prediction for the Volta potential of glycerol, its polar nature may cause it to behave similarly to water. Additionally, glycerol is hygroscopic, suggesting that absorbed water from the ambient environment may cause it to gain charges near the Volta potential for water. 

Conversely, Fluorinert \textcolor{black}{drops acquired significantly more negative charge than either water or glycerol, where $\Delta q/\Delta m\approx$ -12~pC/g. We note the radius of Fluorinert drops ranged from 1.1--1.3 mm, much smaller than water due to their lower surface tension (18 mN/m) and larger density (1.94 g/mL)}. Triboelectrically, that Fluorinert charges negative  against the metal tip (or glass syringe) is unsurprising since fluorocarbons reside at the bottom of triboseries \cite{jimidar2023influence}. Furthermore, we suspect that these higher to charge-to-mass ratios reflect the fact that Flourinert is nonpolar, has no ions, and has an extremely high resistivity (2.3 P$\Omega\cdot$cm, compared to 18.2 M$\Omega\cdot$cm typical of our DI water). Lastly, the Debye length (an important parameter in flow electrification) differs significantly between the fluids used here. For salt water, the Debye length is nanometers or less, whereas, for pure water, it is approximately 1 micron. But for insulating liquids, the Debye length can be millimeters or larger. Theories considering the electronic state in the bulk (i.e. Volta potential) likely do not apply when the Debye length is close to the drop size.

\textcolor{black}{From an applications perspective}, the large potentials acquired by fluids with low conductivity evince the persistent electrostatic hazards present across a number of fields. Beyond industrial settings, however, the frictional charging of dielectric fluids may have important implications for geophysical processes on worlds with exotic potamologies. Saturn's moon Titan, for instance, hosts extensive river and lake systems made not of water, but of liquid methane and ethane \cite{mitri2007hydrocarbon}. While previous work has suggested that the charging of hydrocarbon solids can impact the transport of dust on Titan, whether or not flow electrification impacts the transport of hydrocarbon liquids on Titan remains unexplored. \textcolor{black}{Ultimately, we suggest that more experiments are needed to specifically investigate such insulating liquids, whose equilibrium charge may arise from a different balance of charge transfer mechanisms.} 




\begin{figure}
    \centering
    \includegraphics[width = 1.0\columnwidth]{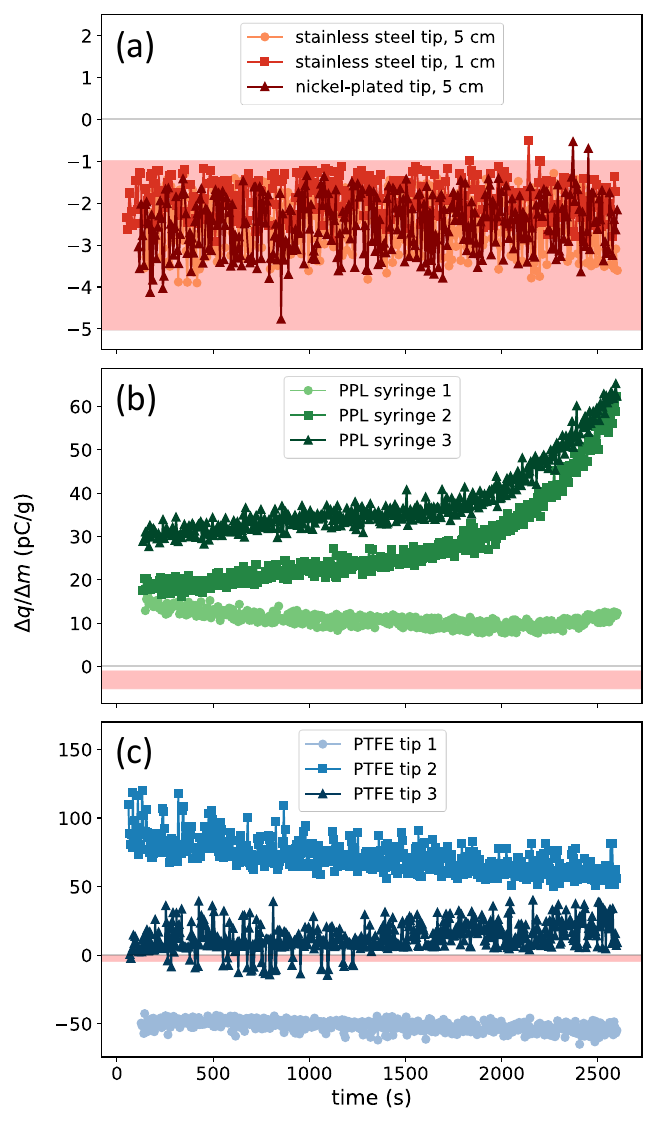}
    \caption{Time series of $\Delta q/\Delta m$ for different syringe and tip components using 15 mL of DI water at a flow rate of 350~$\mu$L/min. The expected $\Delta q / \Delta m$ range of -5~pC/g to -1~pC/g for metal and glass is highlighted by the translucent red region on each plot. Vertical axis limits vary between plots to better depict the magnitude of $\Delta q / \Delta m$. (a) Plot of $\Delta q / \Delta m$ versus time for drops dispensed from glass syringe with different metallic syringe tips. Experiments were performed between 30 and 48\% RH. The tip length varied, yet all tips had a 2.69 mm ID. \textcolor{black}{Across these experiments, $\langle \Delta m \rangle =$ 0.045 $\pm$ 0.002 g.} (b) Plot of $\Delta q / \Delta m$ versus time for drops dispensed from stainless steel tips mounted on PPL syringe bodies.  All tips had a length of 5 cm and 1.70 mm ID, and experiments were performed at 46\% RH. \textcolor{black}{Across these experiments, $\langle \Delta m \rangle =$ 0.0347 $\pm$ 0.0005 g.} (c) Plot of $\Delta q / \Delta m$ versus time for drops dispensed from PTFE tips mounted on a glass syringe body. All tips were cut to a length of 2 cm before Neutrad cleaning and had a 1.35 mm ID, and experiments were performed at $<$ 30\% RH. \textcolor{black}{Across these experiments, $\langle \Delta m \rangle =$ 0.0267 $\pm$ 0.0005 g.} Overall, the introduction of plastic components reveals time and history dependent charging behavior.}
    \label{tipvar_evo}
\end{figure}

\subsection*{Insulating solids}

Instruments for handling of liquids (e.g. pipette tips and syringe bodies) commonly employ plastic components. In solid-solid interactions, plastics can often facilitate the transfer of large amounts of electrostatic charge. Beyond our experiments with glass syringes and metal needles, we also investigated the charge gained by water dispensed from a polypropylene (PPL) syringe body with a 5 cm long stainless steel tip, and water deposited from a glass syringe with a PTFE tip. For these experiments, we used syringes sourced from their original sterile packaging without additional cleaning. Figure \ref{tipvar_evo}a shows that drops dispensed consecutively from a glass syringe/metal tip combination over a period of $\sim$45 minutes have $\langle \Delta q/\Delta m \rangle =$ -2.3 $\pm$ 0.3 pC/g. These time series are consistent with the data presented in Fig.~\ref{heights_pdf}. In contrast, drops sourced from a PPL syringe and metal tip gained positive charge, with $\Delta q/\Delta m =$ 10--30 pC/g (Fig.~\ref{tipvar_evo}b). Moreover, we observed an evolution of $\Delta q/\Delta m$ over long timescales\textcolor{black}{, which is solely due to charge variation since the standard deviation of the drop mass was less than 5\%.} During some trials, $\Delta q/\Delta m$ increased by up to a factor of 2 over the length of the experiment, whereas in other cases we observed a small decrease in $\Delta q/\Delta m$. These experiments show that despite the water passing through 5 cm of metal prior to falling through air, the initial contact with the plastic syringe body dominated the sign, magnitude, and time evolution of the drop charge.

Figure~\ref{tipvar_evo}c shows $\Delta q/\Delta m$ for drops dispensed from a glass syringe and 3 nominally identical PTFE syringe tips (2 cm long). These tips were all cleaned in the same Neutrad bath, rinsed repeatedly with DI water, and stored in a glass beaker until use. They were used sequentially in experiments on the same day with the same water. As in previous experiments, the metallic luer-lock thread of the syringe was grounded. We find that each tip produced drops with different $\Delta q/\Delta m$ magnitudes and polarities. Whereas one tip generated negative drops with $\Delta q / \Delta m$ of -60 pC/g, another generated positive drops with maximum $\Delta q / \Delta m$ of 120 pC/g. Furthermore, as in the experiments with PPL syringes, we observed that $\Delta q/\Delta m$ evolves with time. Our results contrast with those of \citet{BURGO2016}, who report only positive charging behavior for water in contact with PTFE. We note, however, that making a one-to-one comparison between experiments is difficult due to the higher flow rates in \citet{BURGO2016} (140$\times$ greater, which prevented investigation of individual drops). Additionally, their experiments used a plastic reservoir, as in Fig.~\ref{tipvar_evo}b. 


The fact that water drops dispensed from plastic reservoirs or syringe tips gain large amounts of charge is consistent with previous experiments. For example, water drops deposited from a pipette tip can display a large positive charge, $\sim$100~pC or more \cite{choi2013spontaneous}. A more surprising result is the fact that we observe both \textcolor{black}{negative and positive} charging on water droplets interacting with PTFE tips. Like many plastics, PTFE is on the extreme lower end of the triboelectric series and is often considered to be one of the substances that most effectively gains negative charge during frictional interactions. Indeed, the positive electrification of droplets flowing on PTFE surfaces has served as the basis for a number of proposed triboelectric nano generators \cite{kwak2016triboelectrification, wu2021recent}. 

\textcolor{black}{Currently, we do not have} a satisfactory explanation for the evolution of $\Delta q/\Delta m$ when a plastic component is introduced into the system. {\color{black} The magnitude of the charge we observe in experiments with plastics is 10-100 times larger than with glass and metal (Fig.~\ref{heights_pdf}). It is unlikely that an electrochemical charging mechanism similar to Fig.~\ref{water_ions} could produce such a large, and often time-varying charge.} For solid-solid contacts involving hydrocarbons, microscale chemical heterogeneity along solid surfaces can lead to a large variability in charging behavior \cite{grosjean2023asymmetries,sobolev2022charge}. Moreover, history dependence in solid-solid tribocharging has recently been demonstrated during repeated contacts between a sphere and a planar surface using acoustic levitation \cite{grosjean2023single}. In those experiments, hysteresis in water adsorption was suspected to be the primary cause of history dependence. Surprisingly, materials discharged and retested under the same humidity conditions could produce different magnitudes and signs of charge transfer. Similar variability could operate in solid-liquid contacts, explaining the apparent randomness of charge gain we observed in our experiments with plastics. Lastly, surfactants leached from plastic surfaces could also contribute to the diversity of charging behaviors reported here. We suspect that the matter of water-plastic electrification will find a more satisfactory answer in future experiments considering a broader ensemble of plastics.

\section*{Conclusions}

By quasistatically depositing individual drops at low flow rates, we show that, for certain material combinations, charging of water can be attributed to well-characterized electrochemical processes. Specifically, the charge on \textcolor{black}{water} droplets falling from a glass-metal vessel can be described by the Volta potential (Eq.~\ref{volta}). The introduction of plastics, however, can  \textcolor{black}{generate electrification behaviors} which deviate substantially from those predicted by the Volta potential. Indeed, drops interacting with seemingly identical plastic surfaces gained charges that varied drastically both in polarity and magnitude. \textcolor{black}{Furthermore,} the charge \textcolor{black}{magnitude} across subsequent drops could \textcolor{black}{increase} over time, reminiscent of the explosive growth seen in models of triboelectric charging in granular materials \cite{shinbrot2017surface}. \textcolor{black}{Lastly, the involvement of non-polar, non-conductive liquids can also result in highly electrified liquid flow whose behavior warrants targeted studies.}

Together, our results suggest that the Volta potential is but one in a myriad of potential electrification mechanisms leading to flow electrification, and motivate future experiments involving a broader range of solid-fluid (or even fluid-fluid) \textcolor{black}{interactions}. Furthermore, the varied magnitude and polarity of charging we observe with certain material combinations \textcolor{black}{(e.g., water and plastics) undermines the usefulness of tools like triboelectric series.} 
\textcolor{black}{It has been recently shown that an organized triboelectric series can be built by rubbing identical materials together \cite{sobarzo2025spontaneous}, implying that history dependence can be much more important for charging behavior than average, bulk material properties.}
Improved characterizations of  \textcolor{black}{fluid electrification} phenomena will assuredly \textcolor{black}{resolve these open questions, with implications} for industry, the energy sector, and even planetary science. \textcolor{black}{In the meantime, however, our work hints that triboelectric series classification for fluids should be used sparingly and judiciously.}

\section*{Author contributions}
S.M.A. performed all experiments, analyzed the data, and wrote the manuscript. P.E.I. designed and performed preliminary experiments, and wrote the manuscript. J.M.H. formulated the research plan, performed preliminary experiments, and wrote the manuscript. J.C.B. formulated the research plan, analyzed experiments, and wrote the manuscript.

\section*{Conflicts of interest}
There are no conflicts to declare.

\section*{Data availability}



The data analysis scripts for this article are available in an interactive Google Colab notebook at \url{https://colab.research.google.com/drive/1E4XtrUMccv3CFTC_uwi-E-mRX1TQ7zg4?usp=sharing}.

\section*{Acknowledgments}
This work was supported by the Gordon and Betty Moore Foundation, grant DOI 10.37807/gbmf12256. JM was supported by the Portland State ECE startup package.

\bibliography{charged_drops}

\begin{thebibliography}{64}%
\makeatletter
\providecommand \@ifxundefined [1]{%
 \@ifx{#1\undefined}
}%
\providecommand \@ifnum [1]{%
 \ifnum #1\expandafter \@firstoftwo
 \else \expandafter \@secondoftwo
 \fi
}%
\providecommand \@ifx [1]{%
 \ifx #1\expandafter \@firstoftwo
 \else \expandafter \@secondoftwo
 \fi
}%
\providecommand \natexlab [1]{#1}%
\providecommand \enquote  [1]{``#1''}%
\providecommand \bibnamefont  [1]{#1}%
\providecommand \bibfnamefont [1]{#1}%
\providecommand \citenamefont [1]{#1}%
\providecommand \href@noop [0]{\@secondoftwo}%
\providecommand \href [0]{\begingroup \@sanitize@url \@href}%
\providecommand \@href[1]{\@@startlink{#1}\@@href}%
\providecommand \@@href[1]{\endgroup#1\@@endlink}%
\providecommand \@sanitize@url [0]{\catcode `\\12\catcode `\$12\catcode `\&12\catcode `\#12\catcode `\^12\catcode `\_12\catcode `\%12\relax}%
\providecommand \@@startlink[1]{}%
\providecommand \@@endlink[0]{}%
\providecommand \url  [0]{\begingroup\@sanitize@url \@url }%
\providecommand \@url [1]{\endgroup\@href {#1}{\urlprefix }}%
\providecommand \urlprefix  [0]{URL }%
\providecommand \Eprint [0]{\href }%
\providecommand \doibase [0]{http://dx.doi.org/}%
\providecommand \selectlanguage [0]{\@gobble}%
\providecommand \bibinfo  [0]{\@secondoftwo}%
\providecommand \bibfield  [0]{\@secondoftwo}%
\providecommand \translation [1]{[#1]}%
\providecommand \BibitemOpen [0]{}%
\providecommand \bibitemStop [0]{}%
\providecommand \bibitemNoStop [0]{.\EOS\space}%
\providecommand \EOS [0]{\spacefactor3000\relax}%
\providecommand \BibitemShut  [1]{\csname bibitem#1\endcsname}%
\let\auto@bib@innerbib\@empty
\bibitem [{\citenamefont {McNutt}\ and\ \citenamefont {Williams}(2010)}]{mcnutt2010volcanic}%
  \BibitemOpen
  \bibfield  {author} {\bibinfo {author} {\bibfnamefont {S.~R.}\ \bibnamefont {McNutt}}\ and\ \bibinfo {author} {\bibfnamefont {E.~R.}\ \bibnamefont {Williams}},\ }\href@noop {} {\bibfield  {journal} {\bibinfo  {journal} {Bulletin of Volcanology}\ }\textbf {\bibinfo {volume} {72}},\ \bibinfo {pages} {1153} (\bibinfo {year} {2010})}\BibitemShut {NoStop}%
\bibitem [{\citenamefont {M{\'e}ndez~Harper}\ and\ \citenamefont {Dufek}(2016)}]{mendez2016effects}%
  \BibitemOpen
  \bibfield  {author} {\bibinfo {author} {\bibfnamefont {J.}~\bibnamefont {M{\'e}ndez~Harper}}\ and\ \bibinfo {author} {\bibfnamefont {J.}~\bibnamefont {Dufek}},\ }\href@noop {} {\bibfield  {journal} {\bibinfo  {journal} {Journal of Geophysical Research: Atmospheres}\ }\textbf {\bibinfo {volume} {121}},\ \bibinfo {pages} {8209} (\bibinfo {year} {2016})}\BibitemShut {NoStop}%
\bibitem [{\citenamefont {M{\'e}ndez~Harper}\ \emph {et~al.}(2017)\citenamefont {M{\'e}ndez~Harper}, \citenamefont {McDonald}, \citenamefont {Dufek}, \citenamefont {Malaska}, \citenamefont {Burr}, \citenamefont {Hayes}, \citenamefont {McAdams},\ and\ \citenamefont {Wray}}]{mendez2017electrification}%
  \BibitemOpen
  \bibfield  {author} {\bibinfo {author} {\bibfnamefont {J.}~\bibnamefont {M{\'e}ndez~Harper}}, \bibinfo {author} {\bibfnamefont {G.}~\bibnamefont {McDonald}}, \bibinfo {author} {\bibfnamefont {J.}~\bibnamefont {Dufek}}, \bibinfo {author} {\bibfnamefont {M.}~\bibnamefont {Malaska}}, \bibinfo {author} {\bibfnamefont {D.}~\bibnamefont {Burr}}, \bibinfo {author} {\bibfnamefont {A.}~\bibnamefont {Hayes}}, \bibinfo {author} {\bibfnamefont {J.}~\bibnamefont {McAdams}}, \ and\ \bibinfo {author} {\bibfnamefont {J.}~\bibnamefont {Wray}},\ }\href@noop {} {\bibfield  {journal} {\bibinfo  {journal} {Nature Geoscience}\ }\textbf {\bibinfo {volume} {10}},\ \bibinfo {pages} {260} (\bibinfo {year} {2017})}\BibitemShut {NoStop}%
\bibitem [{\citenamefont {M{\'e}ndez~Harper}\ \emph {et~al.}(2021)\citenamefont {M{\'e}ndez~Harper}, \citenamefont {Dufek},\ and\ \citenamefont {McDonald}}]{mendez2021detection}%
  \BibitemOpen
  \bibfield  {author} {\bibinfo {author} {\bibfnamefont {J.}~\bibnamefont {M{\'e}ndez~Harper}}, \bibinfo {author} {\bibfnamefont {J.}~\bibnamefont {Dufek}}, \ and\ \bibinfo {author} {\bibfnamefont {G.~D.}\ \bibnamefont {McDonald}},\ }\href@noop {} {\bibfield  {journal} {\bibinfo  {journal} {Icarus}\ }\textbf {\bibinfo {volume} {357}},\ \bibinfo {pages} {114268} (\bibinfo {year} {2021})}\BibitemShut {NoStop}%
\bibitem [{\citenamefont {Gorman}\ \emph {et~al.}(2024)\citenamefont {Gorman}, \citenamefont {Ruan},\ and\ \citenamefont {Ni}}]{gorman2024electrostatic}%
  \BibitemOpen
  \bibfield  {author} {\bibinfo {author} {\bibfnamefont {M.}~\bibnamefont {Gorman}}, \bibinfo {author} {\bibfnamefont {X.}~\bibnamefont {Ruan}}, \ and\ \bibinfo {author} {\bibfnamefont {R.}~\bibnamefont {Ni}},\ }\href@noop {} {\bibfield  {journal} {\bibinfo  {journal} {Physical Review E}\ }\textbf {\bibinfo {volume} {109}},\ \bibinfo {pages} {034902} (\bibinfo {year} {2024})}\BibitemShut {NoStop}%
\bibitem [{\citenamefont {James}\ \emph {et~al.}(2000)\citenamefont {James}, \citenamefont {Lane},\ and\ \citenamefont {Gilbert}}]{james2000volcanic}%
  \BibitemOpen
  \bibfield  {author} {\bibinfo {author} {\bibfnamefont {M.}~\bibnamefont {James}}, \bibinfo {author} {\bibfnamefont {S.}~\bibnamefont {Lane}}, \ and\ \bibinfo {author} {\bibfnamefont {J.~S.}\ \bibnamefont {Gilbert}},\ }\href@noop {} {\bibfield  {journal} {\bibinfo  {journal} {Journal of Geophysical Research: Solid Earth}\ }\textbf {\bibinfo {volume} {105}},\ \bibinfo {pages} {16641} (\bibinfo {year} {2000})}\BibitemShut {NoStop}%
\bibitem [{\citenamefont {Apodaca}\ \emph {et~al.}(2010)\citenamefont {Apodaca}, \citenamefont {Wesson}, \citenamefont {Bishop}, \citenamefont {Ratner},\ and\ \citenamefont {Grzybowski}}]{apodaca2010contact}%
  \BibitemOpen
  \bibfield  {author} {\bibinfo {author} {\bibfnamefont {M.~M.}\ \bibnamefont {Apodaca}}, \bibinfo {author} {\bibfnamefont {P.~J.}\ \bibnamefont {Wesson}}, \bibinfo {author} {\bibfnamefont {K.~J.}\ \bibnamefont {Bishop}}, \bibinfo {author} {\bibfnamefont {M.~A.}\ \bibnamefont {Ratner}}, \ and\ \bibinfo {author} {\bibfnamefont {B.~A.}\ \bibnamefont {Grzybowski}},\ }\href@noop {} {\bibfield  {journal} {\bibinfo  {journal} {Angew. Chem. Int. Ed}\ }\textbf {\bibinfo {volume} {49}},\ \bibinfo {pages} {946} (\bibinfo {year} {2010})}\BibitemShut {NoStop}%
\bibitem [{\citenamefont {Lacks}\ and\ \citenamefont {Shinbrot}(2019)}]{Lacks2019}%
  \BibitemOpen
  \bibfield  {author} {\bibinfo {author} {\bibfnamefont {D.~J.}\ \bibnamefont {Lacks}}\ and\ \bibinfo {author} {\bibfnamefont {T.}~\bibnamefont {Shinbrot}},\ }\href {\doibase 10.1038/s41570-019-0115-1} {\bibfield  {journal} {\bibinfo  {journal} {Nature Rev. Chemistry}\ }\textbf {\bibinfo {volume} {3}},\ \bibinfo {pages} {465} (\bibinfo {year} {2019})}\BibitemShut {NoStop}%
\bibitem [{\citenamefont {Harris}\ \emph {et~al.}(2019)\citenamefont {Harris}, \citenamefont {Lim},\ and\ \citenamefont {Jaeger}}]{harris2019temperature}%
  \BibitemOpen
  \bibfield  {author} {\bibinfo {author} {\bibfnamefont {I.~A.}\ \bibnamefont {Harris}}, \bibinfo {author} {\bibfnamefont {M.~X.}\ \bibnamefont {Lim}}, \ and\ \bibinfo {author} {\bibfnamefont {H.~M.}\ \bibnamefont {Jaeger}},\ }\href@noop {} {\bibfield  {journal} {\bibinfo  {journal} {Physical Review Materials}\ }\textbf {\bibinfo {volume} {3}},\ \bibinfo {pages} {085603} (\bibinfo {year} {2019})}\BibitemShut {NoStop}%
\bibitem [{\citenamefont {Grosjean}\ and\ \citenamefont {Waitukaitis}(2023{\natexlab{a}})}]{grosjean2023single}%
  \BibitemOpen
  \bibfield  {author} {\bibinfo {author} {\bibfnamefont {G.}~\bibnamefont {Grosjean}}\ and\ \bibinfo {author} {\bibfnamefont {S.}~\bibnamefont {Waitukaitis}},\ }\href@noop {} {\bibfield  {journal} {\bibinfo  {journal} {Physical Review Letters}\ }\textbf {\bibinfo {volume} {130}},\ \bibinfo {pages} {098202} (\bibinfo {year} {2023}{\natexlab{a}})}\BibitemShut {NoStop}%
\bibitem [{\citenamefont {Baytekin}\ \emph {et~al.}(2011)\citenamefont {Baytekin}, \citenamefont {Baytekin}, \citenamefont {Soh},\ and\ \citenamefont {Grzybowski}}]{baytekin2011water}%
  \BibitemOpen
  \bibfield  {author} {\bibinfo {author} {\bibfnamefont {H.~T.}\ \bibnamefont {Baytekin}}, \bibinfo {author} {\bibfnamefont {B.}~\bibnamefont {Baytekin}}, \bibinfo {author} {\bibfnamefont {S.}~\bibnamefont {Soh}}, \ and\ \bibinfo {author} {\bibfnamefont {B.~A.}\ \bibnamefont {Grzybowski}},\ }\href@noop {} {\bibfield  {journal} {\bibinfo  {journal} {Angewandte Chemie (International Edition)}\ }\textbf {\bibinfo {volume} {50}},\ \bibinfo {pages} {0} (\bibinfo {year} {2011})}\BibitemShut {NoStop}%
\bibitem [{\citenamefont {M{\'e}ndez~Harper}\ \emph {et~al.}(2022)\citenamefont {M{\'e}ndez~Harper}, \citenamefont {Harvey}, \citenamefont {Huang}, \citenamefont {McGrath~III}, \citenamefont {Meer},\ and\ \citenamefont {Burton}}]{mendez2022lifetime}%
  \BibitemOpen
  \bibfield  {author} {\bibinfo {author} {\bibfnamefont {J.}~\bibnamefont {M{\'e}ndez~Harper}}, \bibinfo {author} {\bibfnamefont {D.}~\bibnamefont {Harvey}}, \bibinfo {author} {\bibfnamefont {T.}~\bibnamefont {Huang}}, \bibinfo {author} {\bibfnamefont {J.}~\bibnamefont {McGrath~III}}, \bibinfo {author} {\bibfnamefont {D.}~\bibnamefont {Meer}}, \ and\ \bibinfo {author} {\bibfnamefont {J.~C.}\ \bibnamefont {Burton}},\ }\href@noop {} {\bibfield  {journal} {\bibinfo  {journal} {PNAS nexus}\ }\textbf {\bibinfo {volume} {1}},\ \bibinfo {pages} {pgac220} (\bibinfo {year} {2022})}\BibitemShut {NoStop}%
\bibitem [{\citenamefont {M{\'e}ndez~Harper}\ \emph {et~al.}(2024{\natexlab{a}})\citenamefont {M{\'e}ndez~Harper}, \citenamefont {Bumbaugh},\ and\ \citenamefont {Hendon}}]{mendez2024strategies}%
  \BibitemOpen
  \bibfield  {author} {\bibinfo {author} {\bibfnamefont {J.}~\bibnamefont {M{\'e}ndez~Harper}}, \bibinfo {author} {\bibfnamefont {R.~E.}\ \bibnamefont {Bumbaugh}}, \ and\ \bibinfo {author} {\bibfnamefont {C.~H.}\ \bibnamefont {Hendon}},\ }\href@noop {} {\bibfield  {journal} {\bibinfo  {journal} {Iscience}\ }\textbf {\bibinfo {volume} {27}} (\bibinfo {year} {2024}{\natexlab{a}})}\BibitemShut {NoStop}%
\bibitem [{\citenamefont {M{\'e}ndez~Harper}\ \emph {et~al.}(2024{\natexlab{b}})\citenamefont {M{\'e}ndez~Harper}, \citenamefont {McDonald}, \citenamefont {Rheingold}, \citenamefont {Wehn}, \citenamefont {Bumbaugh}, \citenamefont {Cope}, \citenamefont {Lindberg}, \citenamefont {Pham}, \citenamefont {Kim}, \citenamefont {Dufek} \emph {et~al.}}]{mendez2024moisture}%
  \BibitemOpen
  \bibfield  {author} {\bibinfo {author} {\bibfnamefont {J.}~\bibnamefont {M{\'e}ndez~Harper}}, \bibinfo {author} {\bibfnamefont {C.~S.}\ \bibnamefont {McDonald}}, \bibinfo {author} {\bibfnamefont {E.~J.}\ \bibnamefont {Rheingold}}, \bibinfo {author} {\bibfnamefont {L.~C.}\ \bibnamefont {Wehn}}, \bibinfo {author} {\bibfnamefont {R.~E.}\ \bibnamefont {Bumbaugh}}, \bibinfo {author} {\bibfnamefont {E.~J.}\ \bibnamefont {Cope}}, \bibinfo {author} {\bibfnamefont {L.~E.}\ \bibnamefont {Lindberg}}, \bibinfo {author} {\bibfnamefont {J.}~\bibnamefont {Pham}}, \bibinfo {author} {\bibfnamefont {Y.-H.}\ \bibnamefont {Kim}}, \bibinfo {author} {\bibfnamefont {J.}~\bibnamefont {Dufek}},  \emph {et~al.},\ }\href@noop {} {\bibfield  {journal} {\bibinfo  {journal} {Matter}\ }\textbf {\bibinfo {volume} {7}},\ \bibinfo {pages} {266} (\bibinfo {year} {2024}{\natexlab{b}})}\BibitemShut {NoStop}%
\bibitem [{\citenamefont {Gu}\ \emph {et~al.}(2013)\citenamefont {Gu}, \citenamefont {Wei}, \citenamefont {Su},\ and\ \citenamefont {Yu}}]{gu2013role}%
  \BibitemOpen
  \bibfield  {author} {\bibinfo {author} {\bibfnamefont {Z.}~\bibnamefont {Gu}}, \bibinfo {author} {\bibfnamefont {W.}~\bibnamefont {Wei}}, \bibinfo {author} {\bibfnamefont {J.}~\bibnamefont {Su}}, \ and\ \bibinfo {author} {\bibfnamefont {C.~W.}\ \bibnamefont {Yu}},\ }\href@noop {} {\bibfield  {journal} {\bibinfo  {journal} {Scientific reports}\ }\textbf {\bibinfo {volume} {3}},\ \bibinfo {pages} {1337} (\bibinfo {year} {2013})}\BibitemShut {NoStop}%
\bibitem [{\citenamefont {Zhang}\ \emph {et~al.}(2015)\citenamefont {Zhang}, \citenamefont {P{\"a}htz}, \citenamefont {Liu}, \citenamefont {Wang}, \citenamefont {Zhang}, \citenamefont {Shen}, \citenamefont {Ji},\ and\ \citenamefont {Cai}}]{zhang2015electric}%
  \BibitemOpen
  \bibfield  {author} {\bibinfo {author} {\bibfnamefont {Y.}~\bibnamefont {Zhang}}, \bibinfo {author} {\bibfnamefont {T.}~\bibnamefont {P{\"a}htz}}, \bibinfo {author} {\bibfnamefont {Y.}~\bibnamefont {Liu}}, \bibinfo {author} {\bibfnamefont {X.}~\bibnamefont {Wang}}, \bibinfo {author} {\bibfnamefont {R.}~\bibnamefont {Zhang}}, \bibinfo {author} {\bibfnamefont {Y.}~\bibnamefont {Shen}}, \bibinfo {author} {\bibfnamefont {R.}~\bibnamefont {Ji}}, \ and\ \bibinfo {author} {\bibfnamefont {B.}~\bibnamefont {Cai}},\ }\href@noop {} {\bibfield  {journal} {\bibinfo  {journal} {Physical Review X}\ }\textbf {\bibinfo {volume} {5}},\ \bibinfo {pages} {011002} (\bibinfo {year} {2015})}\BibitemShut {NoStop}%
\bibitem [{\citenamefont {Wang}(2021)}]{wang2021contact}%
  \BibitemOpen
  \bibfield  {author} {\bibinfo {author} {\bibfnamefont {Z.~L.}\ \bibnamefont {Wang}},\ }\href@noop {} {\bibfield  {journal} {\bibinfo  {journal} {Reports on Progress in Physics}\ }\textbf {\bibinfo {volume} {84}},\ \bibinfo {pages} {096502} (\bibinfo {year} {2021})}\BibitemShut {NoStop}%
\bibitem [{\citenamefont {Jin}\ \emph {et~al.}(2022)\citenamefont {Jin}, \citenamefont {Wu}, \citenamefont {Sun}, \citenamefont {Wang}, \citenamefont {Cui}, \citenamefont {Zhang},\ and\ \citenamefont {Wang}}]{jin2022electrification}%
  \BibitemOpen
  \bibfield  {author} {\bibinfo {author} {\bibfnamefont {Y.}~\bibnamefont {Jin}}, \bibinfo {author} {\bibfnamefont {C.}~\bibnamefont {Wu}}, \bibinfo {author} {\bibfnamefont {P.}~\bibnamefont {Sun}}, \bibinfo {author} {\bibfnamefont {M.}~\bibnamefont {Wang}}, \bibinfo {author} {\bibfnamefont {M.}~\bibnamefont {Cui}}, \bibinfo {author} {\bibfnamefont {C.}~\bibnamefont {Zhang}}, \ and\ \bibinfo {author} {\bibfnamefont {Z.}~\bibnamefont {Wang}},\ }\href@noop {} {\bibfield  {journal} {\bibinfo  {journal} {Droplet}\ }\textbf {\bibinfo {volume} {1}},\ \bibinfo {pages} {92} (\bibinfo {year} {2022})}\BibitemShut {NoStop}%
\bibitem [{\citenamefont {Armiento}\ \emph {et~al.}(2022)\citenamefont {Armiento}, \citenamefont {Filippeschi}, \citenamefont {Meder},\ and\ \citenamefont {Mazzolai}}]{armiento2022liquid}%
  \BibitemOpen
  \bibfield  {author} {\bibinfo {author} {\bibfnamefont {S.}~\bibnamefont {Armiento}}, \bibinfo {author} {\bibfnamefont {C.}~\bibnamefont {Filippeschi}}, \bibinfo {author} {\bibfnamefont {F.}~\bibnamefont {Meder}}, \ and\ \bibinfo {author} {\bibfnamefont {B.}~\bibnamefont {Mazzolai}},\ }\href@noop {} {\bibfield  {journal} {\bibinfo  {journal} {Communications Materials}\ }\textbf {\bibinfo {volume} {3}},\ \bibinfo {pages} {79} (\bibinfo {year} {2022})}\BibitemShut {NoStop}%
\bibitem [{\citenamefont {Touchard}(2001)}]{TOUCHARD2001}%
  \BibitemOpen
  \bibfield  {author} {\bibinfo {author} {\bibfnamefont {G.}~\bibnamefont {Touchard}},\ }\href {\doibase 10.1016/S0304-3886(01)00081-X} {\bibfield  {journal} {\bibinfo  {journal} {Journal of Electrostatics}\ }\textbf {\bibinfo {volume} {51-52}},\ \bibinfo {pages} {440} (\bibinfo {year} {2001})}\BibitemShut {NoStop}%
\bibitem [{\citenamefont {Rubinstein}\ \emph {et~al.}(2006)\citenamefont {Rubinstein}, \citenamefont {Cohen},\ and\ \citenamefont {Fineberg}}]{rubinstein2006contact}%
  \BibitemOpen
  \bibfield  {author} {\bibinfo {author} {\bibfnamefont {S.~M.}\ \bibnamefont {Rubinstein}}, \bibinfo {author} {\bibfnamefont {G.}~\bibnamefont {Cohen}}, \ and\ \bibinfo {author} {\bibfnamefont {J.}~\bibnamefont {Fineberg}},\ }\href@noop {} {\bibfield  {journal} {\bibinfo  {journal} {Physical review letters}\ }\textbf {\bibinfo {volume} {96}},\ \bibinfo {pages} {256103} (\bibinfo {year} {2006})}\BibitemShut {NoStop}%
\bibitem [{\citenamefont {Li}\ \emph {et~al.}(2021)\citenamefont {Li}, \citenamefont {Chen}, \citenamefont {Yang}, \citenamefont {Zhu}, \citenamefont {Le},\ and\ \citenamefont {Cheng}}]{li2021linear}%
  \BibitemOpen
  \bibfield  {author} {\bibinfo {author} {\bibfnamefont {X.-Y.}\ \bibnamefont {Li}}, \bibinfo {author} {\bibfnamefont {A.}~\bibnamefont {Chen}}, \bibinfo {author} {\bibfnamefont {X.-H.}\ \bibnamefont {Yang}}, \bibinfo {author} {\bibfnamefont {J.-X.}\ \bibnamefont {Zhu}}, \bibinfo {author} {\bibfnamefont {J.-B.}\ \bibnamefont {Le}}, \ and\ \bibinfo {author} {\bibfnamefont {J.}~\bibnamefont {Cheng}},\ }\href@noop {} {\bibfield  {journal} {\bibinfo  {journal} {The Journal of Physical Chemistry Letters}\ }\textbf {\bibinfo {volume} {12}},\ \bibinfo {pages} {7299} (\bibinfo {year} {2021})}\BibitemShut {NoStop}%
\bibitem [{\citenamefont {Klinkenberg}\ and\ \citenamefont {{van der Minne}}(1958)}]{Klinkenberg1958}%
  \BibitemOpen
  \bibfield  {author} {\bibinfo {author} {\bibfnamefont {A.}~\bibnamefont {Klinkenberg}}\ and\ \bibinfo {author} {\bibfnamefont {J.~L.}\ \bibnamefont {{van der Minne}}},\ }\href@noop {} {\emph {\bibinfo {title} {Electrostatics in the Petroleum Industry.}}}\ (\bibinfo  {publisher} {Elsevier Pub. Co.},\ \bibinfo {address} {Amsterdam},\ \bibinfo {year} {1958})\BibitemShut {NoStop}%
\bibitem [{\citenamefont {Solomon}(1959)}]{Solomon1959}%
  \BibitemOpen
  \bibfield  {author} {\bibinfo {author} {\bibfnamefont {T.}~\bibnamefont {Solomon}},\ }\href@noop {} {\emph {\bibinfo {title} {Harmful Effects of Electrostatic Charges on Machinery and Lubricating Oils}}}\ (\bibinfo  {publisher} {Institute of Petroleum},\ \bibinfo {address} {London},\ \bibinfo {year} {1959})\BibitemShut {NoStop}%
\bibitem [{\citenamefont {{ASTM D4865-91}}(1991)}]{ASTM1991}%
  \BibitemOpen
  \bibfield  {author} {\bibinfo {author} {\bibnamefont {{ASTM D4865-91}}},\ }\href@noop {} {\emph {\bibinfo {title} {Standard Guide for Generation and Dissipation of Static Electricity in Petroleum Fuel Systems}}}\ (\bibinfo  {publisher} {American Society for Testing and Materials},\ \bibinfo {year} {1991})\BibitemShut {NoStop}%
\bibitem [{\citenamefont {Harvey}\ \emph {et~al.}(2002)\citenamefont {Harvey}, \citenamefont {Wood}, \citenamefont {Denuault},\ and\ \citenamefont {Powrie}}]{Harvey2002}%
  \BibitemOpen
  \bibfield  {author} {\bibinfo {author} {\bibfnamefont {T.~J.}\ \bibnamefont {Harvey}}, \bibinfo {author} {\bibfnamefont {R.~J.~K.}\ \bibnamefont {Wood}}, \bibinfo {author} {\bibfnamefont {G.}~\bibnamefont {Denuault}}, \ and\ \bibinfo {author} {\bibfnamefont {H.~E.~G.}\ \bibnamefont {Powrie}},\ }\href {\doibase 10.1016/S0301-679X(02)00060-9} {\bibfield  {journal} {\bibinfo  {journal} {Tribology Int.}\ }\textbf {\bibinfo {volume} {35}},\ \bibinfo {pages} {605} (\bibinfo {year} {2002})}\BibitemShut {NoStop}%
\bibitem [{\citenamefont {Mackeown}\ and\ \citenamefont {Wouk}(1942)}]{mackeown1942electrical}%
  \BibitemOpen
  \bibfield  {author} {\bibinfo {author} {\bibfnamefont {S.}~\bibnamefont {Mackeown}}\ and\ \bibinfo {author} {\bibfnamefont {V.}~\bibnamefont {Wouk}},\ }\href@noop {} {\bibfield  {journal} {\bibinfo  {journal} {Industrial \& Engineering Chemistry}\ }\textbf {\bibinfo {volume} {34}},\ \bibinfo {pages} {659} (\bibinfo {year} {1942})}\BibitemShut {NoStop}%
\bibitem [{\citenamefont {Bustin}\ and\ \citenamefont {Dukek}(1983)}]{Bustin1983}%
  \BibitemOpen
  \bibfield  {author} {\bibinfo {author} {\bibfnamefont {W.}~\bibnamefont {Bustin}}\ and\ \bibinfo {author} {\bibfnamefont {W.}~\bibnamefont {Dukek}},\ }\href@noop {} {\emph {\bibinfo {title} {Electrostatics in the Petroleum Industry}}}\ (\bibinfo  {publisher} {Research Studies Press LTD.},\ \bibinfo {address} {Letchworth},\ \bibinfo {year} {1983})\BibitemShut {NoStop}%
\bibitem [{\citenamefont {Hu}\ \emph {et~al.}(2013)\citenamefont {Hu}, \citenamefont {Wang}, \citenamefont {Liu},\ and\ \citenamefont {Gao}}]{Yuqin2013}%
  \BibitemOpen
  \bibfield  {author} {\bibinfo {author} {\bibfnamefont {Y.}~\bibnamefont {Hu}}, \bibinfo {author} {\bibfnamefont {D.}~\bibnamefont {Wang}}, \bibinfo {author} {\bibfnamefont {J.}~\bibnamefont {Liu}}, \ and\ \bibinfo {author} {\bibfnamefont {J.}~\bibnamefont {Gao}},\ }\href {\doibase 10.1088/1742-6596/418/1/012037} {\bibfield  {journal} {\bibinfo  {journal} {J. Phys.: Conf. Ser.}\ }\textbf {\bibinfo {volume} {418}},\ \bibinfo {pages} {012037} (\bibinfo {year} {2013})}\BibitemShut {NoStop}%
\bibitem [{\citenamefont {von Pidoll}\ \emph {et~al.}(1997)\citenamefont {von Pidoll}, \citenamefont {Kr{\"a}mer},\ and\ \citenamefont {Bothe}}]{von1997avoidance}%
  \BibitemOpen
  \bibfield  {author} {\bibinfo {author} {\bibfnamefont {U.}~\bibnamefont {von Pidoll}}, \bibinfo {author} {\bibfnamefont {H.}~\bibnamefont {Kr{\"a}mer}}, \ and\ \bibinfo {author} {\bibfnamefont {H.}~\bibnamefont {Bothe}},\ }\href@noop {} {\bibfield  {journal} {\bibinfo  {journal} {Journal of Electrostatics}\ }\textbf {\bibinfo {volume} {40}},\ \bibinfo {pages} {523} (\bibinfo {year} {1997})}\BibitemShut {NoStop}%
\bibitem [{\citenamefont {Gavis}\ and\ \citenamefont {Wagner}(1968)}]{Gavis1968}%
  \BibitemOpen
  \bibfield  {author} {\bibinfo {author} {\bibfnamefont {J.}~\bibnamefont {Gavis}}\ and\ \bibinfo {author} {\bibfnamefont {J.~P.}\ \bibnamefont {Wagner}},\ }\href {\doibase 10.1016/0009-2509(68)87010-1} {\bibfield  {journal} {\bibinfo  {journal} {Chem. Eng. Sci.}\ }\textbf {\bibinfo {volume} {23}},\ \bibinfo {pages} {381} (\bibinfo {year} {1968})}\BibitemShut {NoStop}%
\bibitem [{\citenamefont {Huber}\ and\ \citenamefont {Sonin}(1977)}]{Huber1977}%
  \BibitemOpen
  \bibfield  {author} {\bibinfo {author} {\bibfnamefont {P.}~\bibnamefont {Huber}}\ and\ \bibinfo {author} {\bibfnamefont {A.}~\bibnamefont {Sonin}},\ }\href {\doibase 10.1016/0021-9797(77)90420-9} {\bibfield  {journal} {\bibinfo  {journal} {J. Colloid Interface Sci.}\ }\textbf {\bibinfo {volume} {61}},\ \bibinfo {pages} {109} (\bibinfo {year} {1977})}\BibitemShut {NoStop}%
\bibitem [{\citenamefont {Bograchev}\ \emph {et~al.}(2012)\citenamefont {Bograchev}, \citenamefont {Martemianov},\ and\ \citenamefont {Paillat}}]{BOGRACHEV2012}%
  \BibitemOpen
  \bibfield  {author} {\bibinfo {author} {\bibfnamefont {D.}~\bibnamefont {Bograchev}}, \bibinfo {author} {\bibfnamefont {S.}~\bibnamefont {Martemianov}}, \ and\ \bibinfo {author} {\bibfnamefont {T.}~\bibnamefont {Paillat}},\ }\href {\doibase 10.1016/j.elstat.2012.07.006} {\bibfield  {journal} {\bibinfo  {journal} {Journal of Electrostatics}\ }\textbf {\bibinfo {volume} {70}},\ \bibinfo {pages} {517} (\bibinfo {year} {2012})}\BibitemShut {NoStop}%
\bibitem [{\citenamefont {Abedian}\ and\ \citenamefont {Sonin}(1982)}]{Abedian1982}%
  \BibitemOpen
  \bibfield  {author} {\bibinfo {author} {\bibfnamefont {B.}~\bibnamefont {Abedian}}\ and\ \bibinfo {author} {\bibfnamefont {A.~A.}\ \bibnamefont {Sonin}},\ }\href {\doibase 10.1017/S0022112082002730} {\bibfield  {journal} {\bibinfo  {journal} {J. Fluid Mech.}\ }\textbf {\bibinfo {volume} {120}},\ \bibinfo {pages} {199} (\bibinfo {year} {1982})}\BibitemShut {NoStop}%
\bibitem [{\citenamefont {Cabaleiro}\ \emph {et~al.}(2008)\citenamefont {Cabaleiro}, \citenamefont {Paillat}, \citenamefont {Moreau},\ and\ \citenamefont {Touchard}}]{CABALEIRO2008}%
  \BibitemOpen
  \bibfield  {author} {\bibinfo {author} {\bibfnamefont {J.~M.}\ \bibnamefont {Cabaleiro}}, \bibinfo {author} {\bibfnamefont {T.}~\bibnamefont {Paillat}}, \bibinfo {author} {\bibfnamefont {O.}~\bibnamefont {Moreau}}, \ and\ \bibinfo {author} {\bibfnamefont {G.}~\bibnamefont {Touchard}},\ }\href {\doibase 10.1016/j.elstat.2007.08.003} {\bibfield  {journal} {\bibinfo  {journal} {Journal of Electrostatics}\ }\textbf {\bibinfo {volume} {66}},\ \bibinfo {pages} {79} (\bibinfo {year} {2008})}\BibitemShut {NoStop}%
\bibitem [{\citenamefont {Stern}(1924)}]{OSTERN1924}%
  \BibitemOpen
  \bibfield  {author} {\bibinfo {author} {\bibfnamefont {O.}~\bibnamefont {Stern}},\ }\href@noop {} {\bibfield  {journal} {\bibinfo  {journal} {Z. Elektrochem.}\ }\textbf {\bibinfo {volume} {30}},\ \bibinfo {pages} {508} (\bibinfo {year} {1924})}\BibitemShut {NoStop}%
\bibitem [{\citenamefont {El-Adawy}\ \emph {et~al.}(2011)\citenamefont {El-Adawy}, \citenamefont {Paillat}, \citenamefont {Cabaleiro},\ and\ \citenamefont {Touchard}}]{eladawy2011}%
  \BibitemOpen
  \bibfield  {author} {\bibinfo {author} {\bibfnamefont {M.}~\bibnamefont {El-Adawy}}, \bibinfo {author} {\bibfnamefont {T.}~\bibnamefont {Paillat}}, \bibinfo {author} {\bibfnamefont {J.~M.}\ \bibnamefont {Cabaleiro}}, \ and\ \bibinfo {author} {\bibfnamefont {G.}~\bibnamefont {Touchard}},\ }\href@noop {} {\bibfield  {journal} {\bibinfo  {journal} {IEEE Trans. Dielectr. Electr. Insul.}\ }\textbf {\bibinfo {volume} {18}},\ \bibinfo {pages} {1463} (\bibinfo {year} {2011})}\BibitemShut {NoStop}%
\bibitem [{\citenamefont {Touchard}\ \emph {et~al.}(1996)\citenamefont {Touchard}, \citenamefont {Patzek},\ and\ \citenamefont {Radke}}]{Touchard1996}%
  \BibitemOpen
  \bibfield  {author} {\bibinfo {author} {\bibfnamefont {G.}~\bibnamefont {Touchard}}, \bibinfo {author} {\bibfnamefont {T.~W.}\ \bibnamefont {Patzek}}, \ and\ \bibinfo {author} {\bibfnamefont {C.~J.}\ \bibnamefont {Radke}},\ }\href@noop {} {\bibfield  {journal} {\bibinfo  {journal} {IEEE Trans. Ind. Appl.}\ }\textbf {\bibinfo {volume} {32}},\ \bibinfo {pages} {1051} (\bibinfo {year} {1996})}\BibitemShut {NoStop}%
\bibitem [{\citenamefont {Washabaugh}\ and\ \citenamefont {Zahn}(1997)}]{Washabaugh1997}%
  \BibitemOpen
  \bibfield  {author} {\bibinfo {author} {\bibfnamefont {A.~P.}\ \bibnamefont {Washabaugh}}\ and\ \bibinfo {author} {\bibfnamefont {M.}~\bibnamefont {Zahn}},\ }\href@noop {} {\bibfield  {journal} {\bibinfo  {journal} {IEEE Trans. Dielectr. Electr. Insul.}\ }\textbf {\bibinfo {volume} {4}},\ \bibinfo {pages} {688} (\bibinfo {year} {1997})}\BibitemShut {NoStop}%
\bibitem [{\citenamefont {Lenard}(1892)}]{lenard1892ueber}%
  \BibitemOpen
  \bibfield  {author} {\bibinfo {author} {\bibfnamefont {P.}~\bibnamefont {Lenard}},\ }\href@noop {} {\bibfield  {journal} {\bibinfo  {journal} {Annalen der Physik}\ }\textbf {\bibinfo {volume} {282}},\ \bibinfo {pages} {584} (\bibinfo {year} {1892})}\BibitemShut {NoStop}%
\bibitem [{\citenamefont {Zhang}\ \emph {et~al.}(2021)\citenamefont {Zhang}, \citenamefont {Chia}, \citenamefont {Fan},\ and\ \citenamefont {Ping}}]{Zhang2021}%
  \BibitemOpen
  \bibfield  {author} {\bibinfo {author} {\bibfnamefont {X.}~\bibnamefont {Zhang}}, \bibinfo {author} {\bibfnamefont {E.}~\bibnamefont {Chia}}, \bibinfo {author} {\bibfnamefont {X.}~\bibnamefont {Fan}}, \ and\ \bibinfo {author} {\bibfnamefont {J.}~\bibnamefont {Ping}},\ }\href {\doibase 10.1038/s41467-021-21974-y} {\bibfield  {journal} {\bibinfo  {journal} {Nature Communications}\ }\textbf {\bibinfo {volume} {12}},\ \bibinfo {pages} {1755} (\bibinfo {year} {2021})}\BibitemShut {NoStop}%
\bibitem [{\citenamefont {Burgo}\ \emph {et~al.}(2016)\citenamefont {Burgo}, \citenamefont {Galembeck},\ and\ \citenamefont {Pollack}}]{BURGO2016}%
  \BibitemOpen
  \bibfield  {author} {\bibinfo {author} {\bibfnamefont {T.~A.~L.}\ \bibnamefont {Burgo}}, \bibinfo {author} {\bibfnamefont {F.}~\bibnamefont {Galembeck}}, \ and\ \bibinfo {author} {\bibfnamefont {G.~H.}\ \bibnamefont {Pollack}},\ }\href {\doibase 10.1016/j.elstat.2016.01.002} {\bibfield  {journal} {\bibinfo  {journal} {Journal of Electrostatics}\ }\textbf {\bibinfo {volume} {80}},\ \bibinfo {pages} {30} (\bibinfo {year} {2016})}\BibitemShut {NoStop}%
\bibitem [{\citenamefont {Kamra}(1982)}]{kamra1982fair}%
  \BibitemOpen
  \bibfield  {author} {\bibinfo {author} {\bibfnamefont {A.}~\bibnamefont {Kamra}},\ }\href@noop {} {\bibfield  {journal} {\bibinfo  {journal} {Journal of Geophysical Research: Oceans}\ }\textbf {\bibinfo {volume} {87}},\ \bibinfo {pages} {4257} (\bibinfo {year} {1982})}\BibitemShut {NoStop}%
\bibitem [{\citenamefont {Hendricks~Jr}(1962)}]{hendricks1962charged}%
  \BibitemOpen
  \bibfield  {author} {\bibinfo {author} {\bibfnamefont {C.~D.}\ \bibnamefont {Hendricks~Jr}},\ }\href@noop {} {\bibfield  {journal} {\bibinfo  {journal} {Journal of Colloid Science}\ }\textbf {\bibinfo {volume} {17}},\ \bibinfo {pages} {249} (\bibinfo {year} {1962})}\BibitemShut {NoStop}%
\bibitem [{\citenamefont {Snarski}\ and\ \citenamefont {Dunn}(1991)}]{snarski1991experiments}%
  \BibitemOpen
  \bibfield  {author} {\bibinfo {author} {\bibfnamefont {S.}~\bibnamefont {Snarski}}\ and\ \bibinfo {author} {\bibfnamefont {P.}~\bibnamefont {Dunn}},\ }\href@noop {} {\bibfield  {journal} {\bibinfo  {journal} {Experiments in Fluids}\ }\textbf {\bibinfo {volume} {11}},\ \bibinfo {pages} {268} (\bibinfo {year} {1991})}\BibitemShut {NoStop}%
\bibitem [{\citenamefont {Mouat}\ \emph {et~al.}(2020)\citenamefont {Mouat}, \citenamefont {Wood}, \citenamefont {Pye},\ and\ \citenamefont {Burton}}]{mouat2020tuning}%
  \BibitemOpen
  \bibfield  {author} {\bibinfo {author} {\bibfnamefont {A.~P.}\ \bibnamefont {Mouat}}, \bibinfo {author} {\bibfnamefont {C.~E.}\ \bibnamefont {Wood}}, \bibinfo {author} {\bibfnamefont {J.~E.}\ \bibnamefont {Pye}}, \ and\ \bibinfo {author} {\bibfnamefont {J.~C.}\ \bibnamefont {Burton}},\ }\href@noop {} {\bibfield  {journal} {\bibinfo  {journal} {Physical Review Letters}\ }\textbf {\bibinfo {volume} {124}},\ \bibinfo {pages} {064502} (\bibinfo {year} {2020})}\BibitemShut {NoStop}%
\bibitem [{\citenamefont {Pye}\ \emph {et~al.}(2018)\citenamefont {Pye}, \citenamefont {Wood},\ and\ \citenamefont {Burton}}]{pye2018precursors}%
  \BibitemOpen
  \bibfield  {author} {\bibinfo {author} {\bibfnamefont {J.~E.}\ \bibnamefont {Pye}}, \bibinfo {author} {\bibfnamefont {C.~E.}\ \bibnamefont {Wood}}, \ and\ \bibinfo {author} {\bibfnamefont {J.~C.}\ \bibnamefont {Burton}},\ }\href@noop {} {\bibfield  {journal} {\bibinfo  {journal} {Physical Review Letters}\ }\textbf {\bibinfo {volume} {121}},\ \bibinfo {pages} {134501} (\bibinfo {year} {2018})}\BibitemShut {NoStop}%
\bibitem [{\citenamefont {Le}\ \emph {et~al.}(2017)\citenamefont {Le}, \citenamefont {Iannuzzi}, \citenamefont {Cuesta},\ and\ \citenamefont {Cheng}}]{le2017PRL}%
  \BibitemOpen
  \bibfield  {author} {\bibinfo {author} {\bibfnamefont {J.}~\bibnamefont {Le}}, \bibinfo {author} {\bibfnamefont {M.}~\bibnamefont {Iannuzzi}}, \bibinfo {author} {\bibfnamefont {A.}~\bibnamefont {Cuesta}}, \ and\ \bibinfo {author} {\bibfnamefont {J.}~\bibnamefont {Cheng}},\ }\href {\doibase 10.1103/PhysRevLett.119.016801} {\bibfield  {journal} {\bibinfo  {journal} {Phys. Rev. Lett.}\ }\textbf {\bibinfo {volume} {119}},\ \bibinfo {pages} {016801} (\bibinfo {year} {2017})}\BibitemShut {NoStop}%
\bibitem [{\citenamefont {Ma}\ \emph {et~al.}(2021)\citenamefont {Ma}, \citenamefont {Xiong}, \citenamefont {Chen},\ and\ \citenamefont {Su}}]{ma2021skpfm}%
  \BibitemOpen
  \bibfield  {author} {\bibinfo {author} {\bibfnamefont {Z.}~\bibnamefont {Ma}}, \bibinfo {author} {\bibfnamefont {X.}~\bibnamefont {Xiong}}, \bibinfo {author} {\bibfnamefont {L.}~\bibnamefont {Chen}}, \ and\ \bibinfo {author} {\bibfnamefont {Y.}~\bibnamefont {Su}},\ }\href@noop {} {\bibfield  {journal} {\bibinfo  {journal} {Electrochimica Acta}\ }\textbf {\bibinfo {volume} {366}},\ \bibinfo {pages} {137422} (\bibinfo {year} {2021})}\BibitemShut {NoStop}%
\bibitem [{\citenamefont {Guo}\ \emph {et~al.}(2012)\citenamefont {Guo}, \citenamefont {Zhao}, \citenamefont {Bai},\ and\ \citenamefont {Qiao}}]{guo2012water}%
  \BibitemOpen
  \bibfield  {author} {\bibinfo {author} {\bibfnamefont {L.}~\bibnamefont {Guo}}, \bibinfo {author} {\bibfnamefont {X.}~\bibnamefont {Zhao}}, \bibinfo {author} {\bibfnamefont {Y.}~\bibnamefont {Bai}}, \ and\ \bibinfo {author} {\bibfnamefont {L.}~\bibnamefont {Qiao}},\ }\href@noop {} {\bibfield  {journal} {\bibinfo  {journal} {Applied Surface Science}\ }\textbf {\bibinfo {volume} {258}},\ \bibinfo {pages} {9087} (\bibinfo {year} {2012})}\BibitemShut {NoStop}%
\bibitem [{\citenamefont {{\"O}rnek}\ \emph {et~al.}(2019)\citenamefont {{\"O}rnek}, \citenamefont {Leygraf},\ and\ \citenamefont {Pan}}]{ornek2019volta}%
  \BibitemOpen
  \bibfield  {author} {\bibinfo {author} {\bibfnamefont {C.}~\bibnamefont {{\"O}rnek}}, \bibinfo {author} {\bibfnamefont {C.}~\bibnamefont {Leygraf}}, \ and\ \bibinfo {author} {\bibfnamefont {J.}~\bibnamefont {Pan}},\ }\href@noop {} {\bibfield  {journal} {\bibinfo  {journal} {Corrosion Engineering, Science and Technology}\ }\textbf {\bibinfo {volume} {54}},\ \bibinfo {pages} {185} (\bibinfo {year} {2019})}\BibitemShut {NoStop}%
\bibitem [{\citenamefont {Trasatti}\ and\ \citenamefont {Lust}(1999)}]{trasatti1999potential}%
  \BibitemOpen
  \bibfield  {author} {\bibinfo {author} {\bibfnamefont {S.}~\bibnamefont {Trasatti}}\ and\ \bibinfo {author} {\bibfnamefont {E.}~\bibnamefont {Lust}},\ }\href@noop {} {\bibfield  {journal} {\bibinfo  {journal} {Modern aspects of electrochemistry}\ ,\ \bibinfo {pages} {1}} (\bibinfo {year} {1999})}\BibitemShut {NoStop}%
\bibitem [{\citenamefont {Mohandas}\ \emph {et~al.}(2024)\citenamefont {Mohandas}, \citenamefont {Bawari}, \citenamefont {Shibuya}, \citenamefont {Ghosh}, \citenamefont {Mondal}, \citenamefont {Narayanan},\ and\ \citenamefont {Cuesta}}]{mohandas2024understanding}%
  \BibitemOpen
  \bibfield  {author} {\bibinfo {author} {\bibfnamefont {N.}~\bibnamefont {Mohandas}}, \bibinfo {author} {\bibfnamefont {S.}~\bibnamefont {Bawari}}, \bibinfo {author} {\bibfnamefont {J.~J.}\ \bibnamefont {Shibuya}}, \bibinfo {author} {\bibfnamefont {S.}~\bibnamefont {Ghosh}}, \bibinfo {author} {\bibfnamefont {J.}~\bibnamefont {Mondal}}, \bibinfo {author} {\bibfnamefont {T.~N.}\ \bibnamefont {Narayanan}}, \ and\ \bibinfo {author} {\bibfnamefont {A.}~\bibnamefont {Cuesta}},\ }\href@noop {} {\bibfield  {journal} {\bibinfo  {journal} {Chemical Science}\ }\textbf {\bibinfo {volume} {15}},\ \bibinfo {pages} {6643} (\bibinfo {year} {2024})}\BibitemShut {NoStop}%
\bibitem [{\citenamefont {Creux}\ \emph {et~al.}(2007)\citenamefont {Creux}, \citenamefont {Lachaise}, \citenamefont {Graciaa},\ and\ \citenamefont {Beattie}}]{creux2007specific}%
  \BibitemOpen
  \bibfield  {author} {\bibinfo {author} {\bibfnamefont {P.}~\bibnamefont {Creux}}, \bibinfo {author} {\bibfnamefont {J.}~\bibnamefont {Lachaise}}, \bibinfo {author} {\bibfnamefont {A.}~\bibnamefont {Graciaa}}, \ and\ \bibinfo {author} {\bibfnamefont {J.~K.}\ \bibnamefont {Beattie}},\ }\href@noop {} {\bibfield  {journal} {\bibinfo  {journal} {The Journal of Physical Chemistry C}\ }\textbf {\bibinfo {volume} {111}},\ \bibinfo {pages} {3753} (\bibinfo {year} {2007})}\BibitemShut {NoStop}%
\bibitem [{\citenamefont {Zimmermann}\ \emph {et~al.}(2001)\citenamefont {Zimmermann}, \citenamefont {Dukhin},\ and\ \citenamefont {Werner}}]{zimmermann2001electrokinetic}%
  \BibitemOpen
  \bibfield  {author} {\bibinfo {author} {\bibfnamefont {R.}~\bibnamefont {Zimmermann}}, \bibinfo {author} {\bibfnamefont {S.}~\bibnamefont {Dukhin}}, \ and\ \bibinfo {author} {\bibfnamefont {C.}~\bibnamefont {Werner}},\ }\href@noop {} {\bibfield  {journal} {\bibinfo  {journal} {The Journal of Physical Chemistry B}\ }\textbf {\bibinfo {volume} {105}},\ \bibinfo {pages} {8544} (\bibinfo {year} {2001})}\BibitemShut {NoStop}%
\bibitem [{\citenamefont {Jimidar}\ \emph {et~al.}(2023)\citenamefont {Jimidar}, \citenamefont {Kwiecinski}, \citenamefont {Roozendaal}, \citenamefont {Kooij}, \citenamefont {Gardeniers}, \citenamefont {Desmet},\ and\ \citenamefont {Sotthewes}}]{jimidar2023influence}%
  \BibitemOpen
  \bibfield  {author} {\bibinfo {author} {\bibfnamefont {I.~S.}\ \bibnamefont {Jimidar}}, \bibinfo {author} {\bibfnamefont {W.}~\bibnamefont {Kwiecinski}}, \bibinfo {author} {\bibfnamefont {G.}~\bibnamefont {Roozendaal}}, \bibinfo {author} {\bibfnamefont {E.~S.}\ \bibnamefont {Kooij}}, \bibinfo {author} {\bibfnamefont {H.~J.}\ \bibnamefont {Gardeniers}}, \bibinfo {author} {\bibfnamefont {G.}~\bibnamefont {Desmet}}, \ and\ \bibinfo {author} {\bibfnamefont {K.}~\bibnamefont {Sotthewes}},\ }\href@noop {} {\bibfield  {journal} {\bibinfo  {journal} {ACS applied materials \& interfaces}\ }\textbf {\bibinfo {volume} {15}},\ \bibinfo {pages} {42004} (\bibinfo {year} {2023})}\BibitemShut {NoStop}%
\bibitem [{\citenamefont {Mitri}\ \emph {et~al.}(2007)\citenamefont {Mitri}, \citenamefont {Showman}, \citenamefont {Lunine},\ and\ \citenamefont {Lorenz}}]{mitri2007hydrocarbon}%
  \BibitemOpen
  \bibfield  {author} {\bibinfo {author} {\bibfnamefont {G.}~\bibnamefont {Mitri}}, \bibinfo {author} {\bibfnamefont {A.~P.}\ \bibnamefont {Showman}}, \bibinfo {author} {\bibfnamefont {J.~I.}\ \bibnamefont {Lunine}}, \ and\ \bibinfo {author} {\bibfnamefont {R.~D.}\ \bibnamefont {Lorenz}},\ }\href@noop {} {\bibfield  {journal} {\bibinfo  {journal} {Icarus}\ }\textbf {\bibinfo {volume} {186}},\ \bibinfo {pages} {385} (\bibinfo {year} {2007})}\BibitemShut {NoStop}%
\bibitem [{\citenamefont {Choi}\ \emph {et~al.}(2013)\citenamefont {Choi}, \citenamefont {Lee}, \citenamefont {Im}, \citenamefont {Kang}, \citenamefont {Lim}, \citenamefont {Kim},\ and\ \citenamefont {Kang}}]{choi2013spontaneous}%
  \BibitemOpen
  \bibfield  {author} {\bibinfo {author} {\bibfnamefont {D.}~\bibnamefont {Choi}}, \bibinfo {author} {\bibfnamefont {H.}~\bibnamefont {Lee}}, \bibinfo {author} {\bibfnamefont {D.~J.}\ \bibnamefont {Im}}, \bibinfo {author} {\bibfnamefont {I.~S.}\ \bibnamefont {Kang}}, \bibinfo {author} {\bibfnamefont {G.}~\bibnamefont {Lim}}, \bibinfo {author} {\bibfnamefont {D.~S.}\ \bibnamefont {Kim}}, \ and\ \bibinfo {author} {\bibfnamefont {K.~H.}\ \bibnamefont {Kang}},\ }\href@noop {} {\bibfield  {journal} {\bibinfo  {journal} {Scientific reports}\ }\textbf {\bibinfo {volume} {3}},\ \bibinfo {pages} {2037} (\bibinfo {year} {2013})}\BibitemShut {NoStop}%
\bibitem [{\citenamefont {Kwak}\ \emph {et~al.}(2016)\citenamefont {Kwak}, \citenamefont {Lin}, \citenamefont {Lee}, \citenamefont {Ryu}, \citenamefont {Kim}, \citenamefont {Zhong}, \citenamefont {Chen},\ and\ \citenamefont {Kim}}]{kwak2016triboelectrification}%
  \BibitemOpen
  \bibfield  {author} {\bibinfo {author} {\bibfnamefont {S.~S.}\ \bibnamefont {Kwak}}, \bibinfo {author} {\bibfnamefont {S.}~\bibnamefont {Lin}}, \bibinfo {author} {\bibfnamefont {J.~H.}\ \bibnamefont {Lee}}, \bibinfo {author} {\bibfnamefont {H.}~\bibnamefont {Ryu}}, \bibinfo {author} {\bibfnamefont {T.~Y.}\ \bibnamefont {Kim}}, \bibinfo {author} {\bibfnamefont {H.}~\bibnamefont {Zhong}}, \bibinfo {author} {\bibfnamefont {H.}~\bibnamefont {Chen}}, \ and\ \bibinfo {author} {\bibfnamefont {S.-W.}\ \bibnamefont {Kim}},\ }\href@noop {} {\bibfield  {journal} {\bibinfo  {journal} {ACS nano}\ }\textbf {\bibinfo {volume} {10}},\ \bibinfo {pages} {7297} (\bibinfo {year} {2016})}\BibitemShut {NoStop}%
\bibitem [{\citenamefont {Wu}\ \emph {et~al.}(2021)\citenamefont {Wu}, \citenamefont {Li}, \citenamefont {Ping},\ and\ \citenamefont {Ying}}]{wu2021recent}%
  \BibitemOpen
  \bibfield  {author} {\bibinfo {author} {\bibfnamefont {X.}~\bibnamefont {Wu}}, \bibinfo {author} {\bibfnamefont {X.}~\bibnamefont {Li}}, \bibinfo {author} {\bibfnamefont {J.}~\bibnamefont {Ping}}, \ and\ \bibinfo {author} {\bibfnamefont {Y.}~\bibnamefont {Ying}},\ }\href@noop {} {\bibfield  {journal} {\bibinfo  {journal} {Nano Energy}\ }\textbf {\bibinfo {volume} {90}},\ \bibinfo {pages} {106592} (\bibinfo {year} {2021})}\BibitemShut {NoStop}%
\bibitem [{\citenamefont {Grosjean}\ and\ \citenamefont {Waitukaitis}(2023{\natexlab{b}})}]{grosjean2023asymmetries}%
  \BibitemOpen
  \bibfield  {author} {\bibinfo {author} {\bibfnamefont {G.}~\bibnamefont {Grosjean}}\ and\ \bibinfo {author} {\bibfnamefont {S.}~\bibnamefont {Waitukaitis}},\ }\href@noop {} {\bibfield  {journal} {\bibinfo  {journal} {Physical Review Materials}\ }\textbf {\bibinfo {volume} {7}},\ \bibinfo {pages} {065601} (\bibinfo {year} {2023}{\natexlab{b}})}\BibitemShut {NoStop}%
\bibitem [{\citenamefont {Sobolev}\ \emph {et~al.}(2022)\citenamefont {Sobolev}, \citenamefont {Adamkiewicz}, \citenamefont {Siek},\ and\ \citenamefont {Grzybowski}}]{sobolev2022charge}%
  \BibitemOpen
  \bibfield  {author} {\bibinfo {author} {\bibfnamefont {Y.~I.}\ \bibnamefont {Sobolev}}, \bibinfo {author} {\bibfnamefont {W.}~\bibnamefont {Adamkiewicz}}, \bibinfo {author} {\bibfnamefont {M.}~\bibnamefont {Siek}}, \ and\ \bibinfo {author} {\bibfnamefont {B.~A.}\ \bibnamefont {Grzybowski}},\ }\href@noop {} {\bibfield  {journal} {\bibinfo  {journal} {Nature Physics}\ }\textbf {\bibinfo {volume} {18}},\ \bibinfo {pages} {1347} (\bibinfo {year} {2022})}\BibitemShut {NoStop}%
\bibitem [{\citenamefont {Shinbrot}\ \emph {et~al.}(2017)\citenamefont {Shinbrot}, \citenamefont {Rutala},\ and\ \citenamefont {Herrmann}}]{shinbrot2017surface}%
  \BibitemOpen
  \bibfield  {author} {\bibinfo {author} {\bibfnamefont {T.}~\bibnamefont {Shinbrot}}, \bibinfo {author} {\bibfnamefont {M.}~\bibnamefont {Rutala}}, \ and\ \bibinfo {author} {\bibfnamefont {H.}~\bibnamefont {Herrmann}},\ }\href@noop {} {\bibfield  {journal} {\bibinfo  {journal} {Physical Review E}\ }\textbf {\bibinfo {volume} {96}},\ \bibinfo {pages} {032912} (\bibinfo {year} {2017})}\BibitemShut {NoStop}%
\bibitem [{\citenamefont {Sobarzo}\ \emph {et~al.}(2025)\citenamefont {Sobarzo}, \citenamefont {Pertl}, \citenamefont {Balazs}, \citenamefont {Costanzo}, \citenamefont {Sauer}, \citenamefont {Foelske}, \citenamefont {Ostermann}, \citenamefont {Pichler}, \citenamefont {Wang}, \citenamefont {Nagata} \emph {et~al.}}]{sobarzo2025spontaneous}%
  \BibitemOpen
  \bibfield  {author} {\bibinfo {author} {\bibfnamefont {J.~C.}\ \bibnamefont {Sobarzo}}, \bibinfo {author} {\bibfnamefont {F.}~\bibnamefont {Pertl}}, \bibinfo {author} {\bibfnamefont {D.~M.}\ \bibnamefont {Balazs}}, \bibinfo {author} {\bibfnamefont {T.}~\bibnamefont {Costanzo}}, \bibinfo {author} {\bibfnamefont {M.}~\bibnamefont {Sauer}}, \bibinfo {author} {\bibfnamefont {A.}~\bibnamefont {Foelske}}, \bibinfo {author} {\bibfnamefont {M.}~\bibnamefont {Ostermann}}, \bibinfo {author} {\bibfnamefont {C.~M.}\ \bibnamefont {Pichler}}, \bibinfo {author} {\bibfnamefont {Y.}~\bibnamefont {Wang}}, \bibinfo {author} {\bibfnamefont {Y.}~\bibnamefont {Nagata}},  \emph {et~al.},\ }\href@noop {} {\bibfield  {journal} {\bibinfo  {journal} {Nature}\ }\textbf {\bibinfo {volume} {638}},\ \bibinfo {pages} {664} (\bibinfo {year} {2025})}\BibitemShut {NoStop}%
\end{thebibliography}%

\end{document}